\documentclass[aps, prl, reprint, amsmath, amssymb, superscriptaddress
]{revtex4-2}
\usepackage[varg]{txfonts}
\usepackage[OT1]{fontenc}
\usepackage[version=4]{mhchem}
\usepackage[normalem]{ulem}
\usepackage{newtxtext,newtxmath,bm,booktabs,array,enumerate,threeparttable,multirow,soul,color,graphicx,amsmath,extarrows,amssymb,dcolumn,natbib,float,tabularx,tensor,xcolor}
\usepackage[colorlinks=true,linkcolor=blue,citecolor=blue,urlcolor=blue]{hyperref} 
\usepackage{orcidlink}
\usepackage{lineno}

\begin{document}

% \linenumbers

%----------------------------------------------------------------------------------------------------
\title{Continuous Gravitational Waves from Thermo-Elastic Mountains in Ultraluminous X-ray Pulsars}
%----------------------------------------------------------------------------------------------------

\newcommand{\LZU}{Lanzhou Center for Theoretical Physics, Key Laboratory of Theoretical Physics of Gansu Province, Key Laboratory of Quantum Theory and Applications of MoE, Gansu Provincial Research Center for Basic Disciplines of Quantum Physics, Lanzhou 730000, China}

\newcommand{\LZUgravity}{Institute of Theoretical Physics \& Research Center of Gravitation, School of Physical Science and Technology, Lanzhou University, Lanzhou 730000, China}

\newcommand{\PKUA}{Department of Astronomy, School of Physics, Peking University, Beijing 100871, China}

\newcommand{\KIAA}{Kavli Institute for Astronomy and Astrophysics, Peking University, Beijing 100871, China}

\newcommand{\NAOC}{National Astronomical Observatories, Chinese Academy of Sciences, Beijing 100101, China}

%---
\author{Hong-Bo Li\orcidlink{0000-0002-4850-8351}}
\email[Contact author: ]{lihb2020@pku.edu.cn}
\affiliation{\LZU}
\affiliation{\LZUgravity}

%---
\author{Ziming Wang\orcidlink{0000-0002-8742-8397}}
\affiliation{\PKUA}
\affiliation{\KIAA}

%---
\author{Lijing Shao\orcidlink{0000-0002-1334-8853}}
\affiliation{\KIAA}
\affiliation{\NAOC}

%---
\author{Ren-Xin Xu\orcidlink{0000-0002-9042-3044}}
\email[Contact author: ]{r.x.xu@pku.edu.cn}
\affiliation{\PKUA}
\affiliation{\KIAA}

%---------------------------------------------------------------------
\begin{abstract}

Fast-spinning, nonaxisymmetrically deformed neutron stars (NSs) are promising targets for continuous gravitational-wave (CGW) searches with current and next-generation ground-based detectors. Ultraluminous X-ray pulsars (ULXPs), characterized by super-Eddington accretion and strong magnetic fields, provide extreme environments in which substantial nonaxisymmetric deformations may develop. In this Letter, we investigate thermo-elastic mountains generated by magnetically induced temperature asymmetries in the accreted crusts of ULXPs. We connect the crustal thermal structure to an accretion-column model and solve the full anisotropic heat-transport equation, extending previous perturbative treatments into the nonlinear, strongly magnetized regime. We find that the weak-field approximation begins to break down near $B_{\rm ref}\sim10^{13}\,{\rm G}$, beyond which nonlinear anisotropic heat transport significantly modifies the quadrupolar thermal response. The resulting thermal asymmetries generate substantial thermo-elastic mass quadrupoles at higher field strengths. Although the currently known ULXPs are unfavorable CGW targets because of their relatively slow spins and, in most cases, large distances, our illustrative spin-evolution calculations show that a NS can enter the super-Eddington accretor phase with a spin period of order $20\,{\rm ms}$. Such rapidly rotating Galactic ULXPs could lie within the projected sensitivities of next-generation detectors such as the Einstein Telescope and Cosmic Explorer. These results extend thermal-mountain physics into the strongly magnetized regime relevant to ULXPs and connect super-Eddington accretion with prospective CGW observations.

\end{abstract}
%---------------------------------------------------------------------

\date{\today}      

\maketitle

%---------------------------------------------------------------------
\textit{Introduction}---Ultraluminous X-ray sources (ULXs) are compact accreting systems with X-ray luminosities exceeding $10^{39}\,{\rm erg\,s^{-1}}$, well above the Eddington luminosity of a typical stellar-mass compact object \cite{Kaaret:2017tcn,2021AstBu..76....6F,King:2023nft}. They are preferentially found in actively star-forming environments and are commonly associated with young X-ray binaries \cite{Roberts:2002ne,Gao:2003sr,Kovlakas:2020wfu}. Although ULXs were traditionally interpreted as accreting stellar-mass or intermediate-mass black holes \cite{Begelman:2006bi,Poutanen:2006uc,Colbert:1999es, 2006ASPC..352..121M,Maccarone:2007dd,Koliopanos:2017aja}, the discovery of coherent X-ray pulsations from M82~X-2 established that at least a fraction of the ULX population contains accreting neutron stars (NSs) \cite{Bachetti:2014qsa}. In these ultraluminous X-ray pulsars (ULXPs), the magnetic field channels the inflowing material toward the magnetic poles, producing localized accretion regions and pulsed X-ray emission. ULXPs therefore provide direct evidence that magnetized NSs can sustain super-Eddington accretion.

The physical conditions that enable such extreme accretion remain uncertain. Population-synthesis calculations suggest that NSs in both high-mass and lower-mass binary systems may pass through ultraluminous phases during episodes of intense mass transfer \cite{King:2001pm, Shao:2015eha, Wiktorowicz:2018duf, Quast:2019csj, Misra:2020oox, 2024A&A...682A..69M}. Models of individual ULXPs span a broad range of inferred magnetic-field strengths. At relatively modest dipole fields, strong outflows and geometrical beaming can substantially enhance the apparent luminosity \cite{Middleton:2016qss, 2017MNRAS.471L..71M, King:2019jvg,Lasota:2023xef}, whereas magnetar-strength fields can reduce the effective scattering opacity and help confine radiation-pressure-dominated accretion columns above the magnetic poles \cite{1976MNRAS.175..395B, Tong:2014esa, DallOsso:2015nzq, Eksi:2014lya, Mushtukov:2015zea, 2020PASJ...72...12I}. Moreover, the magnetic field relevant to crustal heat transport needs not coincide with the large-scale external dipole field inferred from the accretion or spin evolution. Strong non-dipolar or internal magnetic components may also be present \cite{Israel:2016chx,Tsygankov:2018bcz}, providing additional motivation for exploring a broad range of crustal field strengths extending into the strongly magnetized regime. Other processes, including enhanced neutrino cooling \cite{Mushtukov:2018pvq} and photon-bubble instabilities \cite{1992ApJ...388..561A, Begelman:2006ap, 2021MNRAS.508..617Z}, may further modify the accretion dynamics. These extreme magnetic and thermal conditions can, in turn, strongly affect heat transport within the accreted NS crust and motivate our investigation of the nonlinear, strongly magnetized regime.

A rapidly rotating NS with a nonaxisymmetric mass distribution can emit continuous gravitational waves (CGWs). Possible mechanisms for sustaining such deformations include elastic stresses \cite{Haskell:2006sv, Johnson-McDaniel:2012wbj, Gittins:2020cvx, Morales:2022wxs,Morales:2024hwl}, magnetic stresses \cite{Melatos:2005ez, Haskell:2007bh,2001PASA...18..421M, 2004MNRAS.351..569P, Gao:2022hzd, 2025ApJ...979...10R, 2026arXiv260328857H, 2026arXiv260817508B}, and $r$-mode instabilities \cite{Andersson:1998qs,Levin:1998wa, Arras:2002dw, Bondarescu:2008qx}. Accretion provides an additional route to a persistent mass quadrupole. In particular, lateral temperature asymmetries in an accreted crust can shift temperature-sensitive reaction layers and induce pressure and compositional perturbations that are supported by the elastic crust, producing a thermo-elastic ``mountain'' \cite{Bildsten:1998ey, Ushomirsky:2000ax}. Internal magnetic fields provide a natural mechanism for generating such temperature asymmetries because electron heat transport becomes anisotropic when the magnetization parameter $\beta=\omega_{\rm B}\tau$ becomes appreciable \cite{Osborne:2019iph, Hutchins:2022chj}, where $\omega_{\rm B}=eB/(m_e^\ast c)$ is the electron gyrofrequency, $m_e^\ast$ is the effective electron mass, and $\tau$ is the electron relaxation time. Previous calculations of magnetically induced thermal mountains have largely relied on perturbative treatments valid for $\beta \ll 1$. The stronger internal fields relevant to ULXPs can instead drive the crust into the $\beta \gtrsim 1$ regime, where transverse heat transport is strongly suppressed and the weak-field expansion is no longer reliable.

In this Letter, we investigate thermo-elastic deformations in ULXPs generated by magnetically induced temperature asymmetries in the accreted crust. We use a radiation-pressure-dominated accretion-column model to determine the characteristic outer thermal boundary condition of the crust, and then calculate the resulting nonaxisymmetric thermal structure. In the weakly magnetized regime we adopt the perturbative formalism, whereas beyond its range of validity we solve the full multidimensional anisotropic heat-transport equation. We find that the weak-field approximation begins to break down near $B_{\rm ref}\sim10^{13}\,{\rm G}$ for our fiducial crust model, beyond which nonlinear anisotropic heat transport significantly modifies the quadrupolar thermal response and the resulting thermo-elastic mass quadrupole. We further calculate the associated CGW emission and examine whether rapidly rotating Galactic ULXPs could be detectable by current and next-generation ground-based GW observatories, including the Laser Interferometer Gravitational-Wave Observatory \cite[LIGO;][]{KAGRA:2013rdx}, the Einstein Telescope \cite[ET;][]{Branchesi:2023mws}, and Cosmic Explorer \cite[CE;][]{Srivastava:2022slt}. Our calculations therefore connect super-Eddington column accretion, strongly magnetized crustal heat transport, thermo-elastic deformation, and CGW emission within a common framework, and motivate targeted CGW searches for rapidly rotating Galactic ULXPs.

%---------------------------------------------------------------------
\textit{Accretion-column thermal boundary}---At the super-Eddington accretion rates characteristic of ULXPs, radiation pressure decelerates the inflowing gas above the stellar surface and produces an optically thick accretion column \cite{1975PASJ...27..311I, 1975A&A....42..311B, 1976MNRAS.175..395B,Mushtukov:2015zea}. To estimate the thermal condition at the base of the column, we calculate a sequence of representative accretion-column structures with different mass-accretion rates.

The electron and ion temperatures are evolved separately in the accretion-column calculation, but efficient electron--ion energy exchange drives them toward thermal equilibrium near the stellar surface. We therefore characterize the column base by $T_{\rm col,b}\equiv T_e(R_*)\simeq T_i(R_*)$. For the representative super-Eddington accretion rates considered here, $T_{\rm col,b}$ increases with $\dot M$ and spans approximately $(2-3)\times10^8\,{\rm K}$. We use this temperature to set the effective outer boundary of the spherically symmetric crustal background,
%---
\begin{equation}\label{eq:column_background_boundary}
T_0(R_{\rm OB})=T_{\rm col,b}
\simeq(2-3)\times10^8\,{\rm K} \,.
\end{equation}
%---
The accretion-column calculation therefore supplies only the thermal boundary of the background model; its angular structure is not imposed on the crust. The nonaxisymmetric temperature distribution considered below is instead generated internally by magnetically induced anisotropic heat transport. Further details are provided in Sec.~\ref{sec:column_dynamics} of the Supplemental Material.
%---------------------------------------------------------------------

%---------------------------------------------------------------------
\textit{Temperature perturbations induced by magnetic fields}---We next calculate the nonaxisymmetric temperature distribution generated by magnetically anisotropic heat transport in the accreted crust. The unperturbed thermal structure is described by the spherically symmetric temperature $T_0(r)$ and radial heat flux $F_0(r)$ obtained from
%----
\begin{align}
\frac{{\rm d}F_0}{{\rm d}r}
&=
Q_{\rm nuc}(r)-Q_\nu(r,T_0)-\frac{2F_0}{r} \,,
\label{eq:main_background_flux}\\
\frac{{\rm d}T_0}{{\rm d}r}
&=
-\frac{F_0}{\kappa_0}\,.
\label{eq:main_background_temperature}
\end{align}
%----
Here, $Q_{\rm nuc}$ is the volumetric heating rate due to deep-crustal nuclear reactions, $Q_\nu$ is the neutrino-cooling emissivity, and $\kappa_0$ is the unmagnetized electron thermal conductivity. The corresponding microphysical expressions and boundary conditions are given in Sec.~\ref{sec:TS} of the Supplemental Material. Unless otherwise stated, the calculations below use the fiducial crustal background with $\langle\dot M\rangle=\dot M_{\rm Edd} \simeq 1.27\times10^{18}\,{\rm g\,s^{-1}} $, $Q_{\rm imp}=1$, and $T_0(R_{\rm OB})=2.5\times10^8\,{\rm K}$.

The importance of magnetic anisotropy is characterized by the electron magnetization parameter
%----
\begin{equation}\label{eq:main_beta}
\beta
\equiv
\omega_{\rm B}\tau
=
\frac{eB}{m_e^\ast c}\tau \,,
\end{equation}
%----
where $m_e^\ast=p_F/v_F$ and $\tau$ is the electron relaxation time. The limits $\beta\ll1$ and $\beta\gtrsim1$ correspond, respectively, to weakly and appreciably magnetized electron heat transport. In the classical relaxation-time approximation,
%----
\begin{equation}\label{eq:main_conductivity_components}
\kappa_\parallel=\kappa_0\,, \qquad
\kappa_\perp=\frac{\kappa_0}{1+\beta^2}\,,\qquad
\kappa_{\rm H}=\frac{\beta\kappa_0}{1+\beta^2} \,.
\end{equation}
%----
Here, $\kappa_\parallel$ and $\kappa_\perp$ are the electron thermal conductivities parallel and perpendicular to the local magnetic field, respectively, while $\kappa_{\rm H}$ is the Hall thermal-conductivity coefficient. Detailed expressions for the transport coefficients are given in Sec.~\ref{sec:magnetic_solver} of the Supplemental Material.

We prescribe a crust-confined toroidal magnetic field with a quadrupolar $(\ell,m)=(2,2)$ angular structure and characterize its strength by $B_{\rm ref}$ at a reference point near the middle of the crust. The same magnetic geometry is used in both transport regimes. For $\beta\ll1$, we calculate the quadrupolar temperature perturbation using the linear formalism of \citet{Osborne:2019iph}. In this regime, the magnetic driving term is linear in $B_{\rm ref}$ and hence $|\delta T_{22}|\propto B_{\rm ref}$.

Once $\beta$ approaches unity, the magnetic correction to the heat flux can no longer be treated perturbatively. We therefore solve the full multidimensional steady-state equation
%----
\begin{equation}\label{eq:main_strong_heat}
\nabla\cdot
\left[
\boldsymbol{\kappa}(T,\mathbf B)\cdot\nabla T
\right]
+
Q_{\rm nuc}(r)-Q_\nu(r,T)
=0 \,,
\end{equation}
%----
retaining the complete anisotropic conductivity tensor. The angular average of the nonlinear solution is
%----
\begin{equation}
\langle T\rangle(r)
=
\frac{1}{4\pi}
\int T(r,\theta,\phi)\,{\rm d}\Omega \,,
\end{equation}
%----
and we define the non-spherical component by $\delta T=T-\langle T\rangle$. Its signed real $(\ell,m)=(2,2)$ coefficients are obtained by projection onto the same harmonic basis used to specify the magnetic geometry. Details of the field normalization, weak-field equations, harmonic projection, boundary conditions, and numerical implementation are given in Sec.~\ref{sec:magnetic_solver} of the Supplemental Material.

%%%%%%%%%%%%%%%%%%%%%%%%%%%%%%%%%%%%%%%
\begin{figure}[t]
    \centering
    \includegraphics[width=8cm]{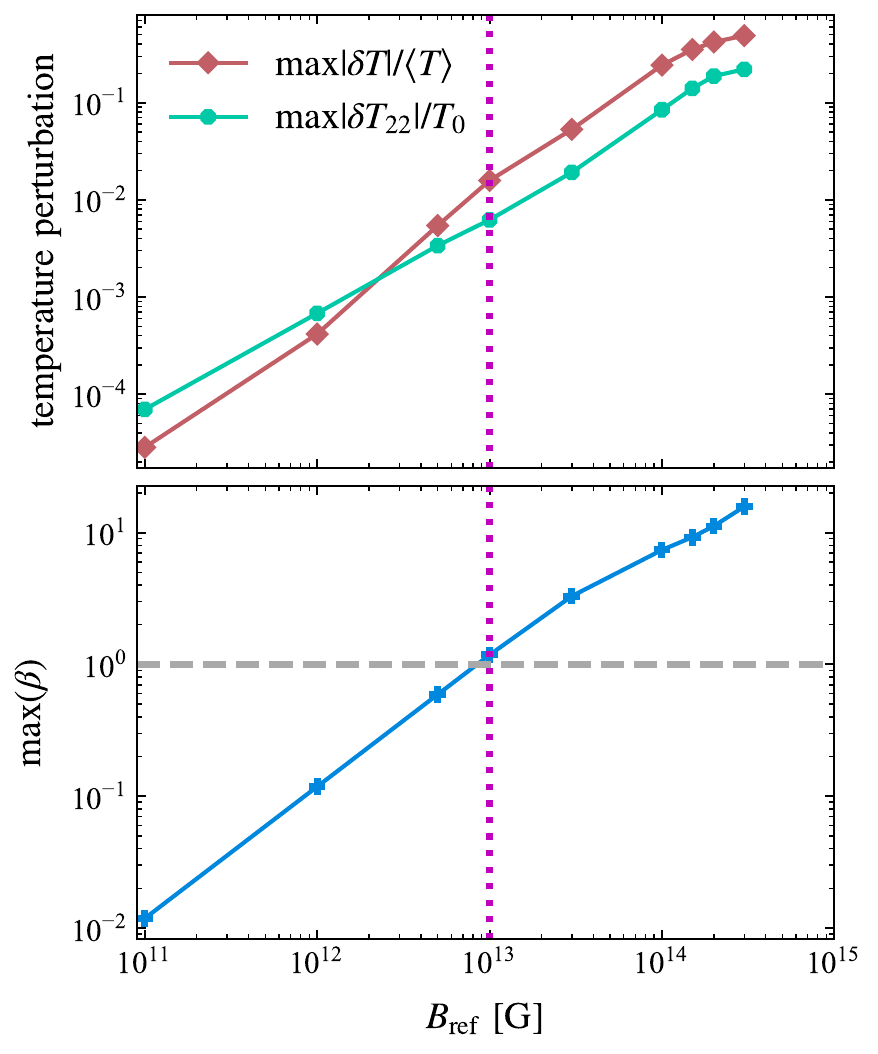}
    \caption{Magnetic-field dependence of the temperature asymmetry and electron magnetization for the fiducial background with $\langle\dot M\rangle=\dot M_{\rm Edd}$, $Q_{\rm imp}=1$, and $T_0(R_{\rm OB})=2.5\times10^8\,{\rm K}$. The upper panel shows the maximum local fractional temperature variation, $\max[|\delta T|/\langle T\rangle]$, and the maximum quadrupolar coefficient, $\max(|\delta T_{22}|/T_0)$. The lower panel shows the maximum magnetization parameter $\max(\beta)$ within the computational crust. The horizontal dashed line denotes $\max(\beta)=1$, while the vertical dotted line marks $B_{\rm ref}=10^{13}\,{\rm G}$, approximately corresponding to the transition between the perturbative and strongly anisotropic transport regimes.}
\label{fig:Fig_Delt_Temperature_2}
\end{figure}
%%%%%%%%%%%%%%%%%%%%%%%%%%%%%%%%%%%%%%%

Figure~\ref{fig:Fig_Delt_Temperature_2} illustrates the transition from perturbative to strongly anisotropic heat transport. At $B_{\rm ref}=10^{12}\,{\rm G}$, the crust remains weakly magnetized, with $\max\beta\simeq0.12$, and the full fixed-field calculation gives $\max(|\delta T_{22}|/T_0)\simeq6.8\times10^{-4}$, in agreement with the weak-field solution at the percent level. The difference remains within approximately $2\%$ at $B_{\rm ref}=5\times10^{12}\,{\rm G}$. A clear departure from the perturbative scaling appears near $B_{\rm ref}\sim10^{13}\,{\rm G}$, where $\max\beta\simeq1.18$. At this field strength, $\max(|\delta T_{22}|/T_0)\simeq6.2\times10^{-3}$, approximately $9.5\%$ below the linear weak-field extrapolation. We therefore identify $B_{\rm ref}\sim10^{13}\,{\rm G}$ as the characteristic onset of nonlinear anisotropic heat transport for the adopted crust model.

At larger magnetic-field strengths, the thermal response can no longer be described as a simple rescaling of the perturbative solution. For $B_{\rm ref}=10^{14}\,{\rm G}$, the maximum local fractional temperature variation and quadrupolar coefficient reach $\max[|\delta T|/\langle T\rangle]\simeq0.24$ and $\max(|\delta T_{22}|/T_0)\simeq8.5\times10^{-2}$, respectively. At the largest field considered here, $B_{\rm ref}=3\times10^{14}\,{\rm G}$, the crust is strongly magnetized, with $\max\beta\simeq15.7$, while the corresponding thermal asymmetries increase to $\max[|\delta T|/\langle T\rangle]\simeq0.49$ and $\max(|\delta T_{22}|/T_0)\simeq0.22$. The rapid growth of the temperature asymmetry reflects the strong suppression of transverse electron heat transport as $\beta$ becomes much larger than unity.

We restrict the quantitative calculations to $B_{\rm ref}\leq3\times10^{14}\,{\rm G}$. At even larger fields, the increasingly extreme transport anisotropy makes the thermal solution more sensitive to angular resolution, while magnetic modifications of the microscopic transport coefficients, including quantization effects, may also become important. The robust result of the present calculation is therefore the transition from the perturbative regime to nonlinear anisotropic heat transport near $B_{\rm ref}\sim10^{13}\,{\rm G}$ and the substantial enhancement of the crustal temperature asymmetry at higher field strengths.

%---------------------------------------------------------------------
\textit{Prospects for future GW detections}---Rapidly rotating ULXPs with nonaxisymmetric deformations can emit CGWs through quadrupole radiation \citep{1979PhRvD..20..351Z, Jaranowski:1998qm}. We characterize the deformation by the stellar ellipticity,
%----
\begin{equation}
\epsilon
=
\sqrt{\frac{8\pi}{15}}
\frac{|Q_{22}|}{I_{\rm zz}} \,,
\end{equation}
%----
where we adopt $I_{\rm zz}=I_0=10^{45}\,{\rm g\,cm^2}$. The corresponding intrinsic GW strain amplitude is \citep{LIGOScientific:2019yhl, Reed:2021scb}
%----
\begin{equation}\label{eq:strain_2}
h_0
=
\frac{4\pi^2G}{c^4}
\frac{I_{\rm zz}\epsilon f_{\rm GW}^2}{d} \,,
\end{equation}
%----
where $f_{\rm GW}=2\nu$ for a NS rotating at spin frequency $\nu$, and $d$ is the source distance.

As discussed by \citet{Ushomirsky:2000ax}, lateral temperature asymmetries in an accreted crust shift temperature-sensitive electron-capture layers and thereby generate elastic deformations. We calculate this response using the signed $(\ell,m)=(2,2)$ temperature coefficients obtained from the thermal calculation. The cosine and sine components are propagated separately through the elastic equations, and the physical quadrupole amplitude is evaluated as $|Q_{22}|=[(Q_{22}^{c})^2+(Q_{22}^{s})^2]^{1/2}$. Details are given in Sec.~\ref{sec:sm_elastic} of the Supplemental Material.

Within the fixed-domain elastic treatment, the calculated quadrupole increases markedly as heat transport becomes strongly anisotropic. We obtain $|Q_{22}|\simeq1.0\times10^{36}\,\mathrm{g\,cm^2}$ at $B_{\rm ref}=10^{12}\,\mathrm{G}$, $8.6\times10^{36}\,\mathrm{g\,cm^2}$ at $10^{13}\,\mathrm{G}$, and $7.8\times10^{37}\,\mathrm{g\,cm^2}$ at $10^{14}\,\mathrm{G}$. At the largest field considered here, $B_{\rm ref}=3\times10^{14}\,\mathrm{G}$, the signed nonlinear temperature components yield
$|Q_{22}|\simeq3.2\times10^{38}\,\mathrm{g\,cm^2}$ and $\epsilon\simeq4.2\times10^{-7}$. We use this largest-field result as an illustrative benchmark for the CGW estimates below. Its sustainability remains to be established because the elastic domain is not updated to account for local melting in the nonlinear thermal solution.

We first evaluate the currently known ULXPs using Eq.~(\ref{eq:strain_2}). Their expected strain amplitudes remain below the adopted sensitivities of LIGO O5, ET, and CE, even for the largest deformation considered here. This is mainly a consequence of their comparatively low spin frequencies and, for most known systems, their extragalactic distances. Detectability therefore depends critically on whether rapidly rotating Galactic ULXPs can exist.

Although the currently known ULXPs are unfavorable CGW targets, the relevant spin frequency is determined by the evolutionary state at which sustained surface accretion begins. We therefore perform an illustrative spin-evolution calculation to examine whether a rapidly rotating NS can enter a super-Eddington accretor phase while retaining a short spin period.

Because the mass quadrupole considered here is generated by accretion-induced temperature asymmetries, we assume that the thermal mountain is absent before sustained surface accretion is established. Accordingly, its GW torque does not affect the preceding ejector or ideal-propeller evolution. Once the system enters the accretor phase, $Q_{22}$ evolves toward the steady-state value obtained from the thermal-transport and elastic-deformation calculations, and the associated GW torque is included thereafter. The torque prescription and the treatment of the quadrupole evolution are described in Sec.~\ref{sec:spin_evolution} of the Supplemental Material.

%%%%%%%%%%%%%%%%%%%%%%%%%%%%%%%%%%%%%%%
\begin{figure}[t]
    \centering
    \includegraphics[width=8cm]{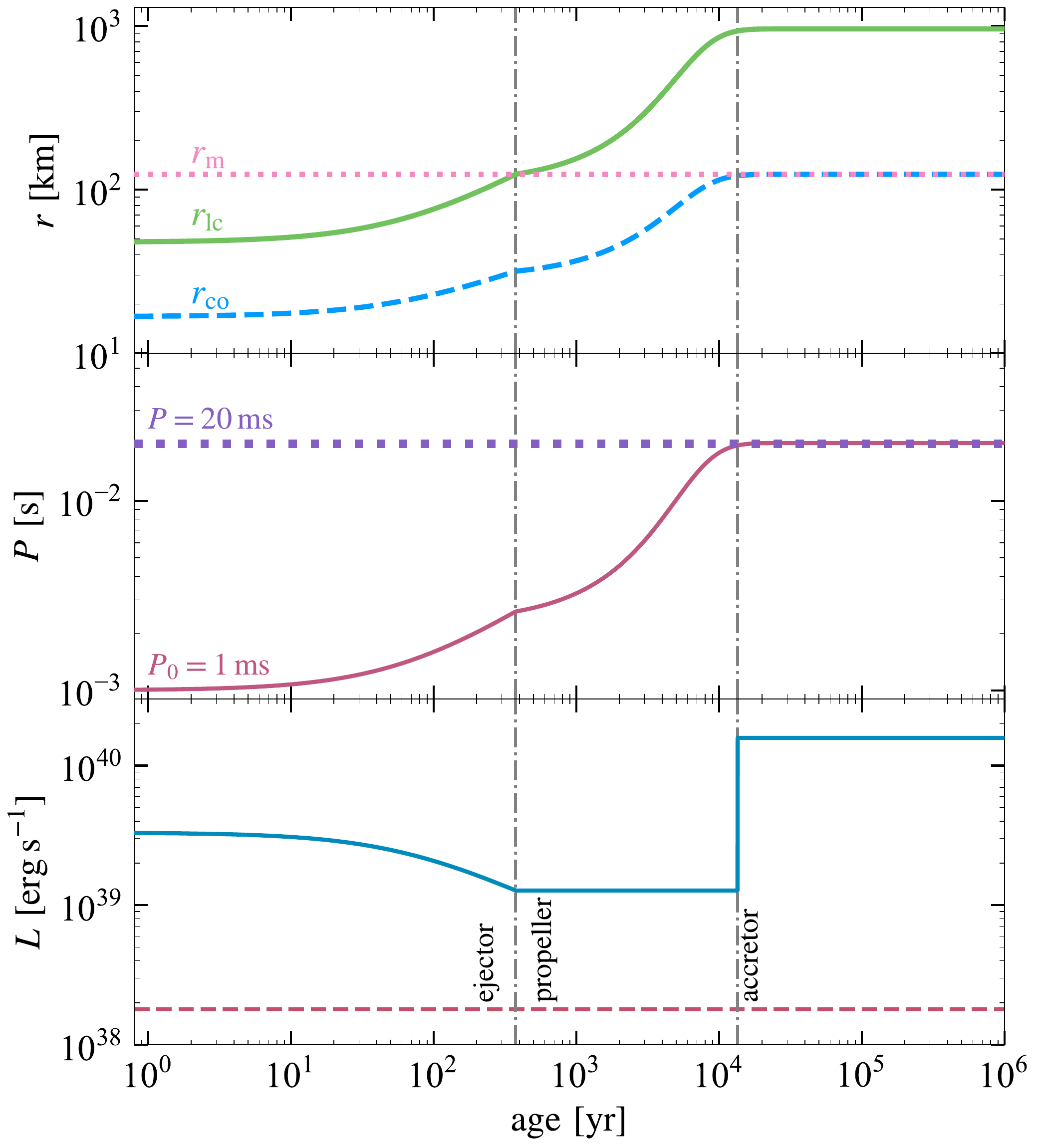}
    \caption{Illustrative evolutionary history leading to the ULXP accretor state. The upper panel shows the light-cylinder radius $r_{\rm lc}$, magnetospheric radius $r_{\rm m}$, and corotation radius $r_{\rm co}$ for an external polar dipole field $B_{\rm p}=10^{12}\,{\rm G}$ and an effective inflow rate $\dot M_{\rm in}=8.5\times10^{19}\,{\rm g\,s^{-1}}$. The vertical lines mark the transitions from the ejector to propeller phase, and from the propeller to the near-equilibrium accretor phase. The middle panel shows the spin-period evolution; the accretion-induced thermal quadrupole develops only after the onset of surface accretion and approaches the fiducial $B_{\rm ref}=3\times10^{14}\,{\rm G}$ steady-state value. The lower panel shows the characteristic gravitational power estimated from $L_{\rm grav}\simeq GM_*\dot M_{\rm in}/r$, with $r=r_{\rm lc}$, $r_{\rm m}$, and $R_*$ in the ejector, propeller, and accretor phases, respectively. The horizontal dashed line denotes the Eddington luminosity. The values shown during the ejector and propeller phases should be regarded as characteristic power scales rather than as ULXP luminosities.}
\label{fig:evolution}
\end{figure}
%%%%%%%%%%%%%%%%%%%%%%%%%%%%%%%%%%%%%%%

Figure~\ref{fig:evolution} follows the system from the ejector phase through the propeller phase and into the accretor state. The preceding phases need not be ultraluminous, since sustained accretion onto the stellar surface has not yet been established. For the adopted parameters, the system enters the super-Eddington accretor regime at $P_{\rm acc}\simeq19.6\,{\rm ms}$, at which point a persistent accretion column can form and the accretion-induced temperature asymmetry, thermo-elastic deformation, and associated mass quadrupole can develop. Although this evolutionary track is not intended to represent a unique history of ULXPs, it demonstrates that super-Eddington surface accretion can coexist with spin periods of order tens of milliseconds for suitable binary and magnetic-field parameters, and hence that rapidly rotating ULXPs are not excluded by the preceding spin evolution.

%%%%%%%%%%%%%%%%%%%%%%%%%%%%%%%%%%%%%%%
\begin{figure}[t]
    \centering
    \includegraphics[width=8cm]{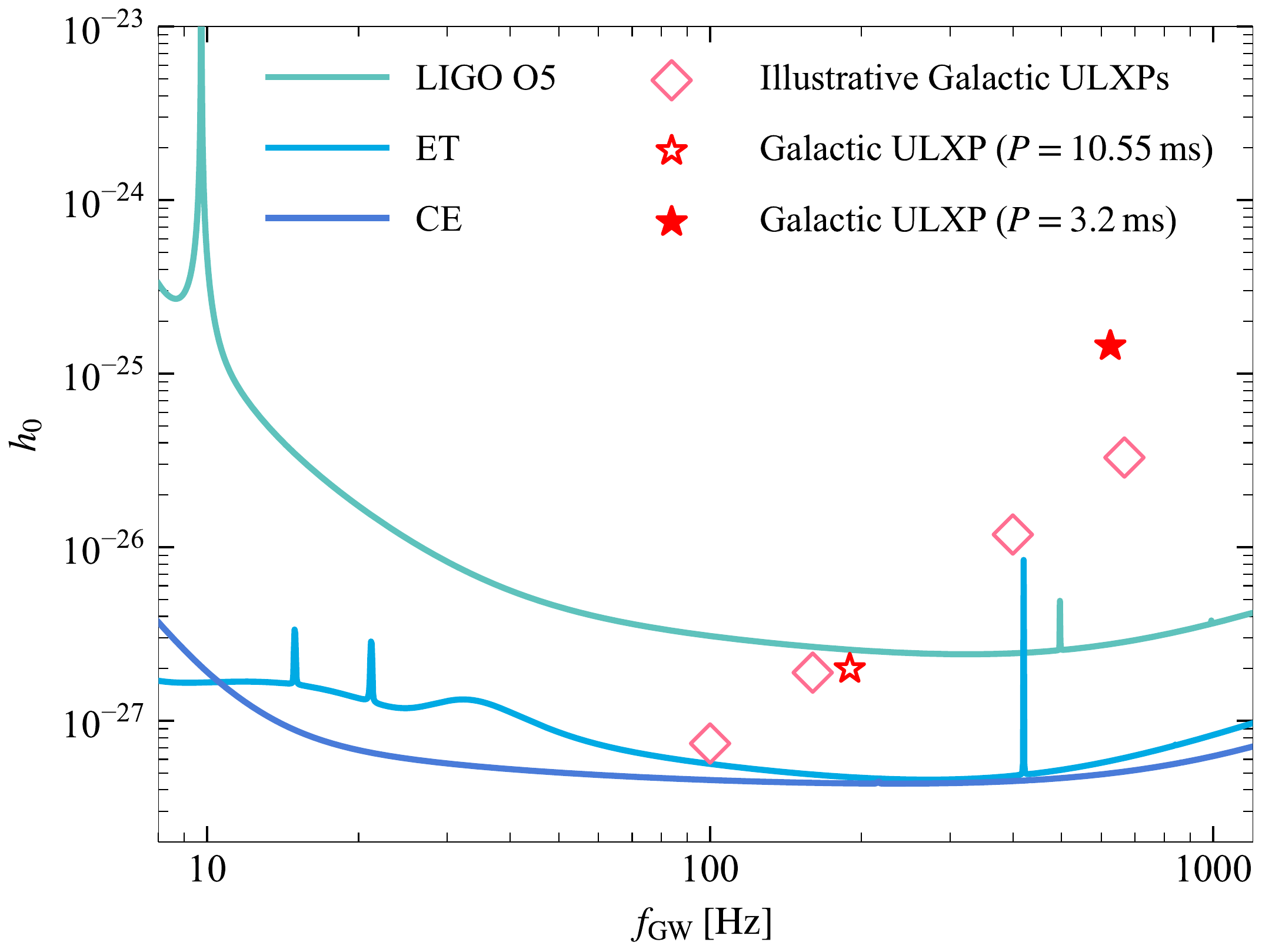}
    \caption{Intrinsic gravitational-wave strain amplitude as a function of  $f_{\rm GW}$. Open diamonds represent illustrative Galactic ULXPs  with spin periods of $P=20$, $12.5$, $5$, and $3\,{\rm ms}$, all  placed at a fiducial distance of $6\,{\rm kpc}$. The open and filled  stars show two additional illustrative Galactic ULXPs with  $P=10.55\,{\rm ms}$ and $P=3.2\,{\rm ms}$, placed at adopted  distances of $8$ and $1.2\,{\rm kpc}$, respectively. All source  markers adopt the $B_{\rm ref}=3\times10^{14}\,{\rm G}$  thermal-mountain model, corresponding to  $\epsilon\simeq4.2\times10^{-7}$. The source markers are intended  to illustrate the strain levels attainable for rapidly rotating  Galactic ULXPs rather than to represent predictions for specific  observed pulsars. For detectors, we use the LIGO O5 strain amplitude spectral density, the combined ET-D sensitivity, and the CE 20-km design sensitivity. The corresponding curves show idealized coherent-search thresholds $h_0^{\rm sens}=11.4\sqrt{S_h(f)/T_{\rm obs}}$ for $T_{\rm obs}=2\,{\rm yr}$, following Ref.~\cite{Watts:2008qw}; no additional detector-network enhancement is applied.}
\label{fig:strain}
\end{figure}
%%%%%%%%%%%%%%%%%%%%%%%%%%%%%%%%%%%%%%%

The strong frequency dependence of Eq.~(\ref{eq:strain_2}) motivates the illustrative Galactic ULXPs shown in Fig.~\ref{fig:strain}. We adopt the $B_{\rm ref}=3\times10^{14}\,{\rm G}$ model, for which $\epsilon \simeq 4.2\times10^{-7}$. Four representative sources with $P=20$, $12.5$, $5$, and $3\,{\rm ms}$ are placed at a fiducial distance of $6\,{\rm kpc}$. We additionally show two rapidly rotating illustrative Galactic ULXPs with $P=10.55$ and $3.2\,{\rm ms}$, at adopted distances of $8$ and $1.2\,{\rm kpc}$, respectively. These additional markers are intended only to illustrate the strain levels attainable for rapidly rotating Galactic ULXPs and do not represent specific observed pulsars. For the sensitivity comparison, we use the LIGO O5 strain amplitude spectral density, the combined ET-D sensitivity, and the CE 20-km design sensitivity. Following Ref.~\cite{Watts:2008qw}, we convert the corresponding amplitude spectral densities $\sqrt{S_h(f)}$ into idealized coherent-search thresholds according to $h_0^{\rm sens}=11.4\sqrt{S_h(f)/T_{\rm obs}}$, with $T_{\rm obs}=2\,{\rm yr}$. No additional detector-network enhancement is applied. Under this assumption, a Galactic ULXP with the deformation considered here becomes accessible at roughly $P \lesssim 22\,{\rm ms}$ for ET and $P \lesssim 25\,{\rm ms}$ for CE, whereas LIGO O5 requires substantially faster rotation, $P \lesssim 11\,{\rm ms}$. These thresholds assume that the signal phase evolution can be modeled accurately enough to maintain coherence over the full two-year observing span. Thus, a Galactic ULXP entering the super-Eddington accretor phase with a spin period of order $20\,{\rm ms}$ could lie within the idealized projected reach of next-generation GW detectors, provided that a thermal deformation near the upper end of the range calculated here can be sustained.

Binary-evolution calculations provide independent motivation for considering a Galactic population of NS ULXs beyond the currently known systems \citep{Shao:2015eha, Misra:2020oox, 2024A&A...682A..69M}. Population-synthesis calculations suggest that NSs can undergo ultraluminous phases during episodes of intense mass transfer, including systems with helium-star donors. These studies motivate considering a broader population of Galactic NS ULXs, while our illustrative spin-evolution calculation shows that rapid rotation can be retained at the onset of the super-Eddington accretor phase for suitable system parameters.

Rapid spin periods comparable to those adopted in Fig.~\ref{fig:strain} are observed among Galactic NSs. For example, PSR J1928$+$1815 has $P=10.55\,{\rm ms}$ \citep{Yang:2025khi}, while PSR J0435$+$3233 has $P\simeq 3.2\,{\rm ms}$ \citep{Wu:2026yky}. More generally, NSs can be recycled to millisecond spin periods through mass transfer and subsequently appear in compact binaries with white-dwarf companions. PSR J1614$-$2230, for instance, has $P \simeq 3\,{\rm ms}$ \citep{2011MNRAS.416.2130T, Tauris:2012jp}. These systems are used here only to motivate the range of rapid spin periods considered in Fig.~\ref{fig:strain}; we do not identify them as ULXPs or post-ULXPs. Together with the illustrative spin evolution shown in Fig.~\ref{fig:evolution}, they demonstrate that the rapid rotation relevant to CGW detectability is physically attainable by NSs. If rapidly rotating Galactic ULXPs exist, the thermo-elastic deformations calculated here could make them promising CGW targets for ET and CE, and potentially for LIGO O5 at the shortest spin periods.

%---------------------------------------------------------------------
\textit{Conclusion}---In this Letter, we investigate thermo-elastic deformations in super-Eddington magnetized NSs undergoing column accretion and their implications for CGW detectability. By coupling a radiation-pressure-dominated accretion-column model to the thermal structure of the accreted crust, we determine the characteristic thermal boundary supplied by the column and calculate the nonaxisymmetric temperature asymmetries generated by magnetically anisotropic heat transport. We find that the weak-field perturbative treatment begins to break down near $B_{\rm ref}\sim10^{13}\,{\rm G}$, beyond which the full multidimensional heat-transport calculation reveals a strongly nonlinear thermal response. The resulting thermo-elastic deformations can produce CGW signals that are potentially detectable from rapidly rotating Galactic ULXPs by next-generation ground-based GW detectors.

The currently known ULXPs remain unfavorable CGW targets because of their relatively slow spins and, in most cases, large distances. However, our illustrative spin-evolution calculation shows that a NS can enter the super-Eddington accretor phase while retaining a spin period of order $20\,{\rm ms}$. For a Galactic ULXP with a thermal deformation near the upper end of the range calculated here, such a spin period would place the source within the projected two-year coherent sensitivities of ET and CE, while LIGO O5 would require substantially faster rotation. Our results therefore extend thermal-mountain calculations into the strongly magnetized, nonlinear transport regime relevant to ULXPs and motivate both future self-consistent magneto-thermal-elastic modeling and targeted CGW searches for rapidly rotating Galactic ULXPs.

%---------------------------------------------------------------------
\textit{Acknowledgments}---This work was supported by the National Natural Science Foundation of China (Grant No. 12447148, 12247101), the Fundamental Research Funds for the Central Universities (Grant No. lzujbky-2026-jdzx02), the Natural Science Foundation of Gansu Province (No. 25JRRA799), the ‘111 Center’ under Grant No. B20063, and the National SKA Program of China (2020SKA0120100, 2020SKA0120300).
%---------------------------------------------------------------------

\bibliography{refs}% Produces the bibliography via BibTeX.

%---------------------------------------------------------------------

\clearpage

%%%%%%%%%% Merge with supplemental materials %%%%%%%%%%
\onecolumngrid
\let\oldsection\section
\renewcommand{\section}{%
\setcounter{equation}{0}
\setcounter{figure}{0}
\setcounter{table}{0}
\oldsection%
}
\setcounter{section}{0}
\setcounter{secnumdepth}{2}
\setcounter{page}{1}
\renewcommand{\theequation}{\thesection\arabic{equation}}
\renewcommand{\thefigure}{\thesection\arabic{figure}}
\renewcommand{\thesection}{\Alph{section}}
\renewcommand{\theHequation}{SM.\thesection.\arabic{equation}}
\renewcommand{\theHfigure}{SM.\thesection.\arabic{figure}}
\renewcommand{\thesubsection}{\thesection.\arabic{subsection}}
\renewcommand{\thesubsubsection}{\thesubsection.\arabic{subsubsection}}
\renewcommand{\thepage}{S\arabic{page}}
\makeatletter

%---------------------------------------------------------------------
\begin{center}
  \textbf{\large Supplemental Material\\[.2cm] 
  Continuous Gravitational Waves from Thermo-Elastic Mountains in Ultraluminous X-ray Pulsars}\\[0.2cm]
  Hong-Bo Li, Ziming Wang, Lijing Shao, and Ren-Xin Xu
\end{center}
%---------------------------------------------------------------------

This Supplemental Material provides further details of the physical model and numerical calculations. In Sec.~\ref{sec:column_dynamics}, we present the governing hydrodynamic equations for the accretion column and the associated boundary conditions. In Sec.~\ref{sec:TS}, we describe the steady-state thermal structure of the spherically symmetric NS background. Based on this background solution, Sec.~\ref{sec:magnetic_solver} details the calculation of magnetically induced temperature asymmetries in the weakly and strongly magnetized regimes. In Sec.~\ref{sec:sm_elastic}, we describe the calculation of the resulting elastic deformation and mass quadrupole moment. Finally, in Sec.~\ref{sec:spin_evolution}, we present the spin-evolution model used to assess the detectability of the associated continuous gravitational waves.

%---------------------------------------------------------------------
\section{Accretion-column thermal boundary}
\label{sec:column_dynamics}
%---------------------------------------------------------------------

The purpose of the accretion-column calculation is to determine the characteristic thermal condition supplied by the super-Eddington accretion flow to the outer boundary of the neutron-star crust. Rather than modelling individual ULXP sources, we construct a sequence of representative accretion-column solutions parameterized by the mass-accretion rate. This allows us to isolate the systematic dependence of the column structure and, in particular, its base temperature on $\dot M$.

We follow the two-fluid radiation-hydrodynamical framework of Ref.~\cite{West:2016xbz}, in which the electron and ion energy equations are evolved separately. We introduce the dimensionless radius, flow velocity, sound speeds, and total energy transport rate,
%----
\begin{equation}
\begin{aligned}
\tilde r &\equiv \frac{r}{R_g},\qquad
\tilde v \equiv \frac{v}{c},\qquad
\tilde a_e \equiv \frac{a_e}{c}\,,\\
\tilde a_i &\equiv \frac{a_i}{c}\,,\qquad
\tilde a_{\rm rad} \equiv \frac{a_{\rm rad}}{c}\,,\qquad
\tilde{\mathcal{E}}\equiv \frac{\dot E}{\dot M c^2}\,,
\end{aligned}
\label{eq:sm_column_variables}
\end{equation}
%----
where $R_g=GM_*/c^2$ and the stellar surface is located at $\tilde r_{\rm sur}=R_*/R_g$. Throughout the accretion-column calculation, we adopt the fiducial scales $M_*=1.4\,M_{\odot}$ and $R_*=10\,{\rm km}$ to convert between the dimensionless and physical variables. The principal quantity carried from the column calculation into the crustal thermal model is the resulting column-base temperature $T_{\rm col,b}$.

The coupled dimensionless conservation equations are
%----
\begin{subequations}
\label{eq:sm_column_odes}
\begin{align}
\frac{d\tilde a_{\rm rad}}{d\tilde r}
&=
\frac{\tilde a_{\rm rad}}{2}
\left(
\frac{3}{\tilde r}
+\frac{1}{\tilde v}
\frac{d\tilde v}{d\tilde r}
\right)
-\frac{\sigma_\parallel R_g}{2m_pc}
\frac{\dot M}{A}
\frac{\gamma_{\rm rad}}{\tilde a_{\rm rad}}
\left(
\tilde{\mathcal{E}}
+\frac{\tilde v^2}{2}
+\frac{\tilde a_e^2}{\gamma_e-1}
+\frac{\tilde a_i^2}{\gamma_i-1}
+\frac{\tilde a_{\rm rad}^2}{\gamma_{\rm rad}-1}
-\frac{1}{\tilde r}
\right)\,,
\label{eq:sm_radiation_ode}\\
\frac{d\tilde a_i}{d\tilde r}
&=
\frac{(1-\gamma_i)\tilde a_i}{2}
\left[
\frac{3}{\tilde r}
+\frac{1}{\tilde v}
\frac{d\tilde v}{d\tilde r}
+\frac{R_g}{c^2}\frac{A}{\dot M}
\frac{\gamma_i\dot U_i}{\tilde a_i^2}
\right]\,,
\label{eq:sm_ion_ode}\\
\frac{d\tilde a_e}{d\tilde r}
&=
\frac{(1-\gamma_e)\tilde a_e}{2}
\left[
\frac{3}{\tilde r}
+\frac{1}{\tilde v}
\frac{d\tilde v}{d\tilde r}
+\frac{R_g}{c^2}\frac{A}{\dot M}
\frac{\gamma_e\dot U_e}{\tilde a_e^2}
\right]\,,
\label{eq:sm_electron_ode}\\
\frac{d\tilde{\mathcal{E}}}{d\tilde r}
&=
\frac{\tilde a_{\rm rad}^2}
{c(\ell_2-\ell_1)
\gamma_{\rm rad}(\gamma_{\rm rad}-1)\tilde v}
\left(
\frac{R_*^3}{R_g\tilde r^3}
\right)^{1/2}
\min\left(
c,\frac{c}{\tau_\perp}
\right)\,,
\label{eq:sm_energy_ode}\\
\frac{d\tilde v}{d\tilde r}
&=
\frac{\tilde v}
{\tilde v^2-\tilde a_i^2-\tilde a_e^2}
\bigg[
\frac{3(\tilde a_i^2+\tilde a_e^2)}{\tilde r}
-\frac{1}{\tilde r^2}
+\frac{\sigma_\parallel R_g}{m_pc}
\frac{\dot M}{A}
\left(
\tilde{\mathcal{E}}
+\frac{\tilde v^2}{2}
+\frac{\tilde a_i^2}{\gamma_i-1}
+\frac{\tilde a_e^2}{\gamma_e-1}
+\frac{\tilde a_{\rm rad}^2}{\gamma_{\rm rad}-1}
-\frac{1}{\tilde r}
\right)
\nonumber\\
&\qquad
+\frac{R_g}{c^2}\frac{A}{\dot M}
\left[
(\gamma_i-1)\dot U_i
+(\gamma_e-1)\dot U_e
\right]
\bigg]\,.
\label{eq:sm_velocity_ode}
\end{align}
\end{subequations}
%----
Here, $A(r)$ is the dipolar accretion-channel cross-sectional area, $\ell_1$ and $\ell_2$ specify the transverse boundaries of the column, and $\tau_\perp$ is the optical depth across the column. The quantities $\dot U_i$ and $\dot U_e$ denote the net ion and electron energy-exchange rates, respectively.

At the upper boundary, the accreting gas approaches free fall,
%----
\begin{equation}
\label{eq:sm_column_top_velocity}
\tilde v_{\rm top}
=
-\left(
\frac{2}{\tilde r_{\rm top}}
\right)^{1/2}\,,
\qquad
\left.
\frac{d\tilde v}{d\tilde r}
\right|_{\rm top}
=
-\left.
\frac{1}{\tilde r^2\tilde v}
\right|_{\rm top}\,.
\end{equation}
%----
The radiation and ion Mach numbers determine the incident radiation and gas temperatures. We impose $T_{i,\rm top}=T_{e,\rm top}$, while electron--ion energy exchange allows the two temperatures to evolve consistently throughout the column.

At the stellar surface, the net radial photon energy flux is set to zero,
%----
\begin{equation}
\label{eq:sm_surface_flux}
F_{\rm rad}(R_*)
=
-\frac{c}{3n_e\sigma_\parallel}
\left.
\frac{dU_{\rm rad}}{dr}
\right|_{R_*}
+\frac{4}{3}
v(R_*)U_{\rm rad}(R_*)
=
0 \,,
\end{equation}
%----
and the total energy transport rate approaches
%----
\begin{equation}
\label{eq:sm_surface_energy}
\dot E(R_*)
=
\frac{GM_*\dot M}{R_*}\,.
\end{equation}
%----
The mass-accretion rate is related to the X-ray luminosity through
%----
\begin{equation}
\label{eq:sm_mdot_lx}
\dot M
=
\frac{L_XR_*}{GM_*} \,.
\end{equation}
%----

The accretion-column calculation follows the two-fluid
radiation-hydrodynamical framework of Ref.~\cite{West:2016xbz}.
The column geometry, magnetic-field-dependent scattering cross
sections, transverse photon escape, and electron--ion energy exchange
are treated within the same physical prescription. We adopt the
fiducial stellar parameters $M_*=1.4\,M_{\odot}$ and
$R_*=10\,{\rm km}$. Since the present calculation is intended to
characterize the general super-Eddington regime rather than individual
ULXPs, we consider three representative mass-accretion rates,
\begin{equation}
\dot M =
3.8\times10^{19},\quad
9.5\times10^{19},\quad
1.9\times10^{20}\ {\rm g\,s^{-1}}.
\end{equation}
The remaining model ingredients are kept within the same fiducial
column prescription, allowing us to isolate the systematic dependence
of the column structure on $\dot M$.

%%%%%%%%%%%%%%%%%%%%%%%%%%%%%%%%%%%%%%%
\begin{figure*}
    \centering
    \includegraphics[width=15cm]{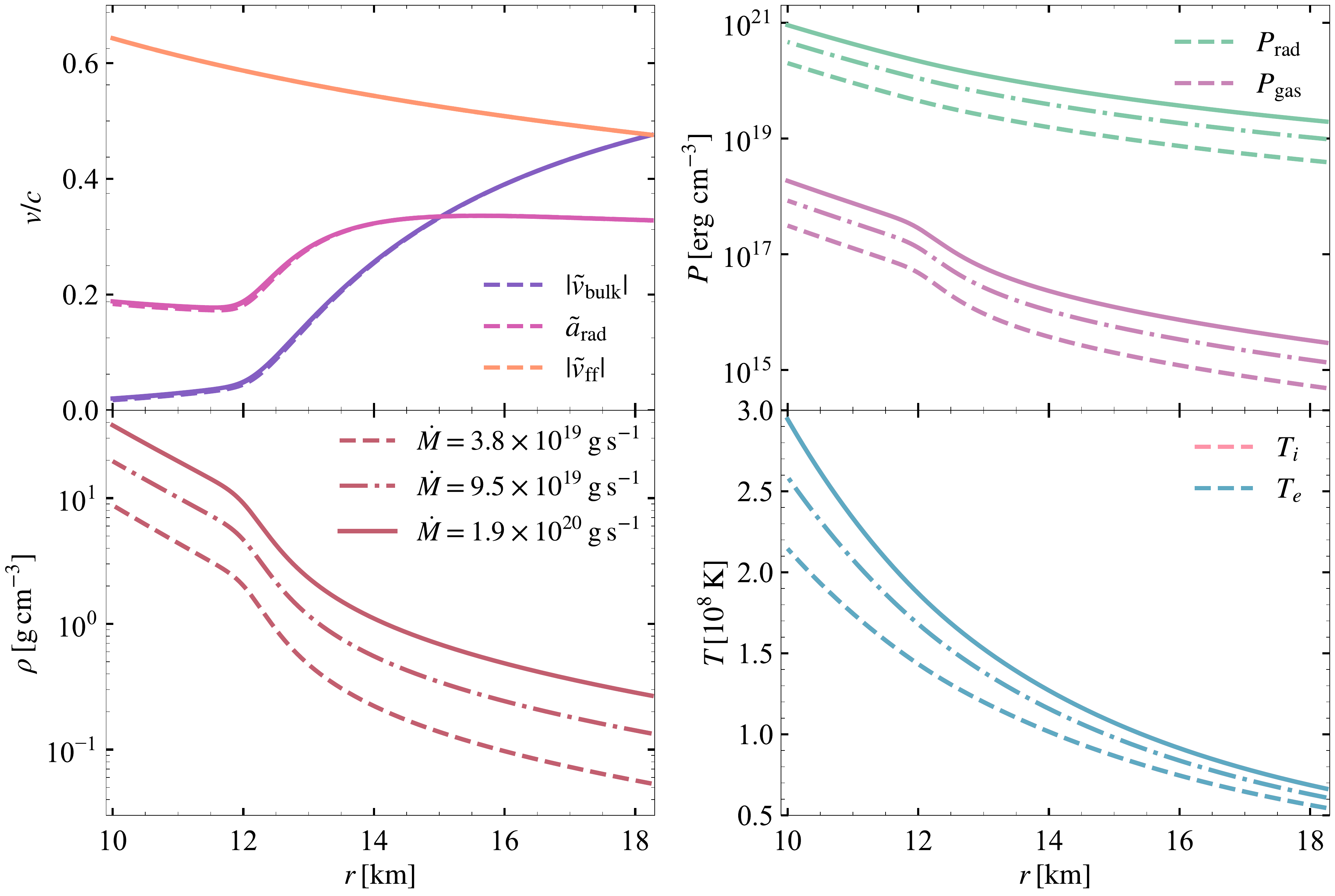}
    \caption{
    Radial profiles of representative accretion-column solutions for
    $\dot M=3.8\times10^{19}$,
    $9.5\times10^{19}$, and
    $1.9\times10^{20}\ {\rm g\,s^{-1}}$.
    The upper-left panel shows the magnitudes of the bulk-flow velocity,
    radiation sound speed, and free-fall velocity; the lower-left panel
    shows the mass density. The upper-right panel shows the radiation
    and gas pressures, and the lower-right panel shows the electron and
    ion temperatures. The radial coordinate is measured from the stellar
    centre, with $r=R_*=10\,{\rm km}$ corresponding to the stellar
    surface. Dashed, dash-dotted, and solid curves denote increasing
    mass-accretion rates, respectively.
    }
    \label{fig:accretion_column}
\end{figure*}
%%%%%%%%%%%%%%%%%%%%%%%%%%%%%%%%%%%%%%%

Figure~\ref{fig:accretion_column} shows that the three solutions share
the same qualitative dynamical structure. The inflowing gas enters the
upper part of the column at a velocity close to free fall and is
progressively decelerated by radiation pressure toward the stellar
surface, with radiation pressure dominating over gas pressure through
most of the column. Increasing $\dot M$ primarily raises the density
and pressure of the flow, while the overall velocity structure changes
more modestly.

The thermal structure also varies systematically with the accretion
rate. Both the electron and ion temperatures increase toward the
stellar surface, and the characteristic temperature of the inner column
rises with increasing $\dot M$. Although the electron and ion energy
equations are evolved separately, efficient Coulomb coupling in the
dense inner column drives the two species toward thermal equilibrium,
so that $T_e\simeq T_i$ near the stellar surface.

We therefore define the characteristic column-base temperature as
\begin{equation}
\label{eq:sm_column_base_temperature}
T_{\rm col,b}\equiv T_e(R_*)\,,
\end{equation}
with the numerical solutions satisfying
$T_e(R_*)\simeq T_i(R_*)$. For the accretion rates considered here,
$T_{\rm col,b}$ increases from approximately
$2\times10^8\,{\rm K}$ to $3\times10^8\,{\rm K}$. We therefore adopt
\begin{equation}
\label{eq:sm_column_temperature_range}
T_{\rm col,b}
\simeq
(2-3)\times10^8\,{\rm K} \,,
\end{equation}
as the characteristic thermal-boundary range for the subsequent crustal
calculation.

The accretion-column calculation is used only to set the effective
outer thermal boundary of the spherically symmetric crustal background,
\begin{equation}
\label{eq:sm_outer_temperature}
T_0(R_{\rm OB})=T_{\rm col,b} \,,
\end{equation}
where $R_{\rm OB}$ denotes the outer boundary of the crustal thermal
domain. In this way, the dependence of the column-base temperature on
$\dot M$ provides a direct connection between the accretion flow and
the thermal state of the underlying crust. The angular structure of the
column is not imposed on the crust; the nonaxisymmetric temperature
perturbations considered below are generated internally by
magnetically induced anisotropic heat transport.

The physical state of the crust is assessed from the spherical background temperature $T_0(r)$ using the Coulomb coupling parameter. Regions identified as liquid in this background are excluded from the elastic calculation. The resulting elastic domain is retained when the nonlinear temperature field is coupled to the elastic equations.

%---------------------------------------------------------------------
\section{Background thermal structure}
\label{sec:TS}
%---------------------------------------------------------------------

The spherically symmetric thermal background is constructed using the DH core equation of state \cite{Douchin:2001sv} together with the HZ accreted-crust composition \cite{1990A&A...229..117H,1990A&A...227..431H}. For the stellar model adopted in the crustal thermal calculation, we use $M\simeq1.84\,M_{\odot}$ and $R\simeq1.17\times10^{6}\,{\rm cm}$. The computational crust extends from $R_{\rm IB}\simeq10.90\,{\rm km}$ to $R_{\rm OB}\simeq11.64\,{\rm km}$. These radii correspond to the radial range covered by the adopted accreted-crust model and define the domain used in both the spherically symmetric background and the multidimensional magnetic heat-transport calculations. The steady radial heat flux $F_0(r)$ and temperature $T_0(r)$ obey
%----
\begin{align}
\frac{{\rm d}F_0}{{\rm d}r}
&=
Q_{\rm nuc}-Q_\nu-\frac{2F_0}{r} \,,
\label{eq:dFdr}
\\
\frac{{\rm d}T_0}{{\rm d}r}
&=
-\frac{F_0}{\kappa_0} \,.
\label{eq:dTdr}
\end{align}
%----

The accretion rate entering the steady crustal heating calculation is interpreted as a long-term average, $\langle\dot M\rangle$, and need not coincide with the instantaneous super-Eddington rate used in the accretion-column calculation. The latter determines the characteristic column-base temperature $T_{\rm col,b}$, whereas $\langle\dot M\rangle$ controls the rate of nuclear energy deposition in the accreted crust.

The nuclear heating rate in the $i$th smeared reaction layer is
%----
\begin{equation}\label{eq:sm_nuclear_heating}
Q_{{\rm nuc},i}
= \frac{\langle\dot M\rangle}{m_b}
\frac{E_{{\rm nuc},i}}
{(4\pi/3)(r_i^3-r_{i+1}^3)} \,,
\end{equation}
%----
where $m_b$ is the baryon mass and $E_{{\rm nuc},i}$ is the heat released per accreted baryon. We adopt the HZ reaction sequence, for which the total heat release is approximately $1.45\,{\rm MeV}$ per accreted baryon. In particular, we distinguish between the accretion rate $\dot M$ used in the accretion-column calculation and the long-term time-averaged accretion rate $\langle\dot M\rangle$ relevant to the steady crustal thermal structure. The former determines the column-base temperature $T_{\rm col,b}$, whereas the latter sets the rate of nuclear energy deposition in the accreted crust. In general, these two accretion rates need not be identical.

The dominant crustal neutrino-cooling process included here is electron--ion bremsstrahlung,
%----
\begin{equation}
Q_\nu
=
3.229\times10^{17}
\rho_{12}T_9^6
\frac{Z^2}{A}(1-X_n)
\ {\rm erg\,s^{-1}\,cm^{-3}} \,,
\end{equation}
%----
where $\rho_{12}=\rho/(10^{12}\,{\rm g\,cm^{-3}})$ and $T_9=T_0/(10^9\,{\rm K})$.

The unmagnetized electron thermal conductivity is
%----
\begin{equation}
\kappa_0
=
\frac{\pi^2}{3}
\frac{k_B^2T_0 n_e}{m_e^\ast}\tau \,,
\end{equation}
%----
with
%----
\begin{equation}
\tau^{-1}
=
\nu_{\rm ep}+\nu_{\rm eQ} \,,
\end{equation}
%----
where
%----
\begin{equation}
\nu_{\rm ep}
=
\frac{13e^2k_BT_0}{\hbar^2c}\,,
\qquad
\nu_{\rm eQ}
=
\frac{4\pi Q_{\rm imp}e^4n_{\rm ion}}
{p_F^2v_F}\Lambda_{\rm imp} \,.
\end{equation}
%----
We set $\Lambda_{\rm imp}=1$ in the numerical calculations.

At the inner boundary, the heat flux is matched to the modified-Urca luminosity of an isothermal core,
%----
\begin{equation}
F_0(R_{\rm IB})
=
-\frac{L_{\rm core}}
{4\pi R_{\rm IB}^2},
\end{equation}
%----
while the outer boundary is set by the accretion-column calculation,
%----
\begin{equation}
T_0(R_{\rm OB})=T_{\rm col,b}.
\end{equation}
%----
We adopt $T_{\rm col,b}=2.5\times10^8\,{\rm K}$ as the fiducial boundary temperature and have verified that varying it over $(2-3)\times10^8\,{\rm K}$ has only a minor effect on the temperature of the deep crust relevant to the thermal-mountain calculation.

For the background models used below, we consider $\langle\dot M\rangle/\dot M_{\rm Edd}=1$, $2$, and $4$. As expected, increasing the long-term accretion rate produces a hotter crust through enhanced deep-crustal heating. We also monitor the local Coulomb coupling parameter,
%----
\begin{equation}
\Gamma_{\rm Coul}
= \frac{Z^2e^2}{a_i k_BT_0} \,,
\qquad
a_i=
\left(
\frac{3}{4\pi n_{\rm ion}}
\right)^{1/3} \,,
\end{equation}
%----
and adopt the canonical one-component-plasma melting criterion $\Gamma_{\rm m}=175$. The deep inner crust remains crystalline for all background models considered here, even when some shallower layers approach or cross the melting threshold at the highest accretion rate. These background solutions provide the temperature, thermal conductivity, and scattering times used in the magnetic heat-transport calculations described in the following section.

%---------------------------------------------------------------------
\section{Magnetization diagnostics and fixed-field thermal solver}
\label{sec:magnetic_solver}
%---------------------------------------------------------------------

We now characterize the magnetic modification of heat transport using the steady crustal backgrounds constructed in Sec.~\ref{sec:TS}. Accordingly, the accretion-rate dependence in this section is parameterized by the long-term time-averaged rate $\langle\dot M\rangle$ introduced above.

The influence of the magnetic field on crustal heat transport is controlled by the electron magnetization parameter
%----
\begin{equation}
\label{eq:sm_beta_explicit}
\beta
\equiv
\omega_{\rm B}\tau
=
\frac{eB}{m_e^\ast c}
\left[
\frac{13e^2k_BT_0}{\hbar^2c}
+
\frac{4\pi Q_{\rm imp}e^4n_{\rm ion}}
{p_F^2v_F}
\Lambda_{\rm imp}
\right]^{-1} \,,
\end{equation}
%----
where $T_0(r)$ and $\tau(r)$ are obtained from the spherical thermal backgrounds described in Sec.~\ref{sec:TS}. Throughout the numerical calculations we set $\Lambda_{\rm imp}=1$.
The relativistic effective electron mass is
%----
\begin{equation}
\label{eq:effective_electron_mass}
m_e^\ast
=
\frac{p_F}{v_F}
=
\left[
m_e^2+
\left(\frac{p_F}{c}\right)^2
\right]^{1/2} \,.
\end{equation}
%----
Toward the deeper crust, the electron density and Fermi momentum increase, leading to a larger $m_e^\ast$ and hence a smaller gyrofrequency $\omega_{\rm B}=eB/(m_e^\ast c)$ at fixed $B$. At the same time, the relaxation time varies with temperature, composition, and impurity scattering. These combined effects generally make $\beta$ smaller in the inner crust and larger toward the outer part of the computational domain. Small-scale structures in the radial profiles arise from the layered HZ composition \cite{1990A&A...229..117H, 1990A&A...227..431H}, in which the nuclear mass and charge numbers change discretely between successive reaction layers.

%%%%%%%%%%%%%%%%%%%%%%%%%%%%%%%%%%%
\begin{figure*}[t]
    \centering
    \includegraphics[width=0.95\textwidth]{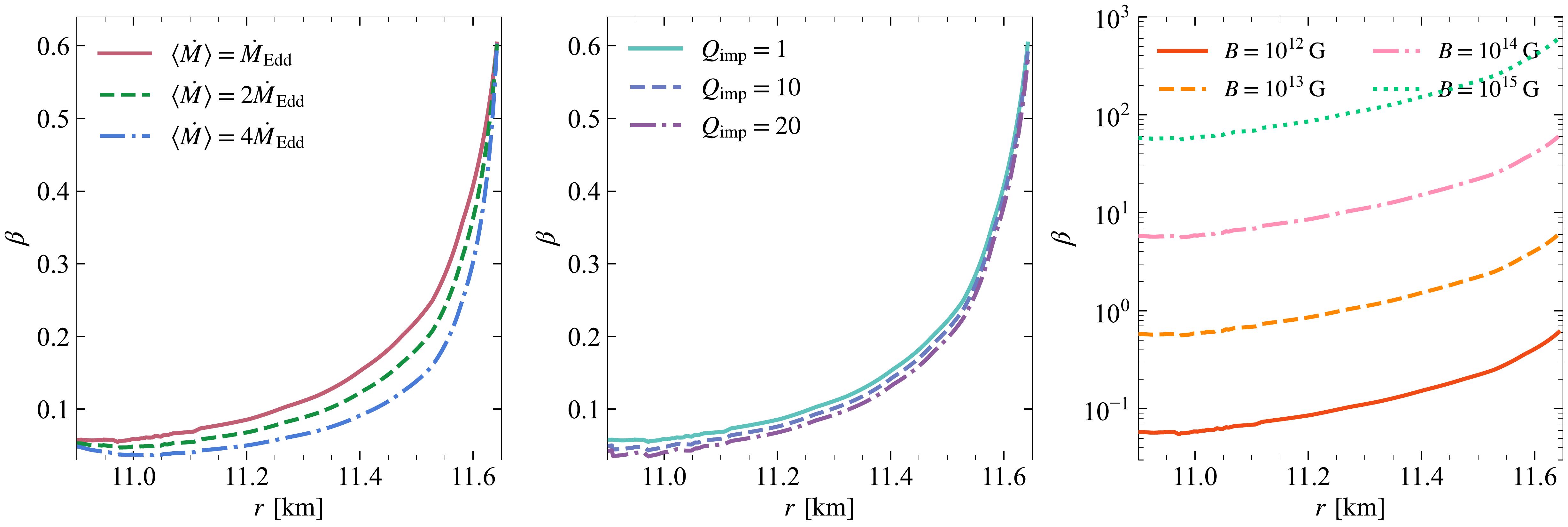}
    \caption{ Radial profiles of the electron magnetization parameter $\beta=\omega_{\rm B}\tau$ in the accreted crust. The left panel compares thermal backgrounds with $\langle\dot M\rangle/\dot M_{\rm Edd}=1$, $2$, and $4$ for $Q_{\rm imp}=1$ and $B=10^{12}\,{\rm G}$. The middle panel shows the dependence on $Q_{\rm imp}=1$, $10$, and $20$ for $\langle\dot M\rangle=\dot M_{\rm Edd}$ and $B=10^{12}\,{\rm G}$. The right panel shows the magnetic-field dependence for $B=10^{12}$, $10^{13}$, $10^{14}$, and $10^{15}\,{\rm G}$ using the fiducial background with $\langle\dot M\rangle=\dot M_{\rm Edd}$ and $Q_{\rm imp}=1$. All thermal backgrounds adopt $T_0(R_{\rm OB})=2.5\times10^8\,{\rm K}$. }
\label{fig:mag_parameter_1}
\end{figure*}
%%%%%%%%%%%%%%%%%%%%%%%%%%%%%%%%%%%

Figure~\ref{fig:mag_parameter_1} shows how the magnetization changes with the long-term accretion rate, impurity content, and magnetic-field strength. Increasing $\langle\dot M\rangle$ raises the crustal temperature through stronger deep-crustal heating. The enhanced temperature-dependent electron scattering shortens $\tau$ and therefore reduces $\beta$. This effect is most pronounced in the deeper crust, while the profiles converge toward the outer boundary because all background models share the same imposed value of $T_0(R_{\rm OB})$. Increasing $Q_{\rm imp}$ also reduces $\beta$ by strengthening electron--impurity scattering. The effect is again stronger in the deeper layers, but remains subdominant compared with the direct dependence on magnetic-field strength.

Within the classical relaxation-time approximation adopted here, $\tau$ has no explicit dependence on $B$, and therefore $\beta\propto B$ at a fixed thermodynamic background. For the fiducial model with $\langle\dot M\rangle=\dot M_{\rm Edd}$ and $Q_{\rm imp}=1$, we obtain
%----
\begin{align*}
\beta &\simeq 0.055\text{--}0.60 \,,
\quad B=10^{12}\,{\rm G}\,,\\
\beta &\simeq 0.55\text{--}6.0 \,,
\quad B=10^{13}\,{\rm G} \,, \\
\beta &\simeq 5.5\text{--}60 \,,
\quad B=10^{14}\,{\rm G}\,, \\
\beta &\simeq 55\text{--}600 \,,
\quad B=10^{15}\, {\rm G} \,.
\end{align*}
%----
Thus, the entire computational crust remains in the weakly magnetized regime at $B=10^{12}\,{\rm G}$, whereas $\beta$ crosses unity within the crust near $B\sim10^{13}\,{\rm G}$. This marks the transition to appreciably anisotropic heat transport. For $B\gtrsim10^{14}\,{\rm G}$, $\beta\gg1$ throughout the computational domain and transverse heat conduction is strongly suppressed.

%%%%%%%%%%%%%%%%%%%%%%%%%%%%%%%%%%%
\begin{figure*}[t]
    \centering
    \includegraphics[width=0.95\textwidth]{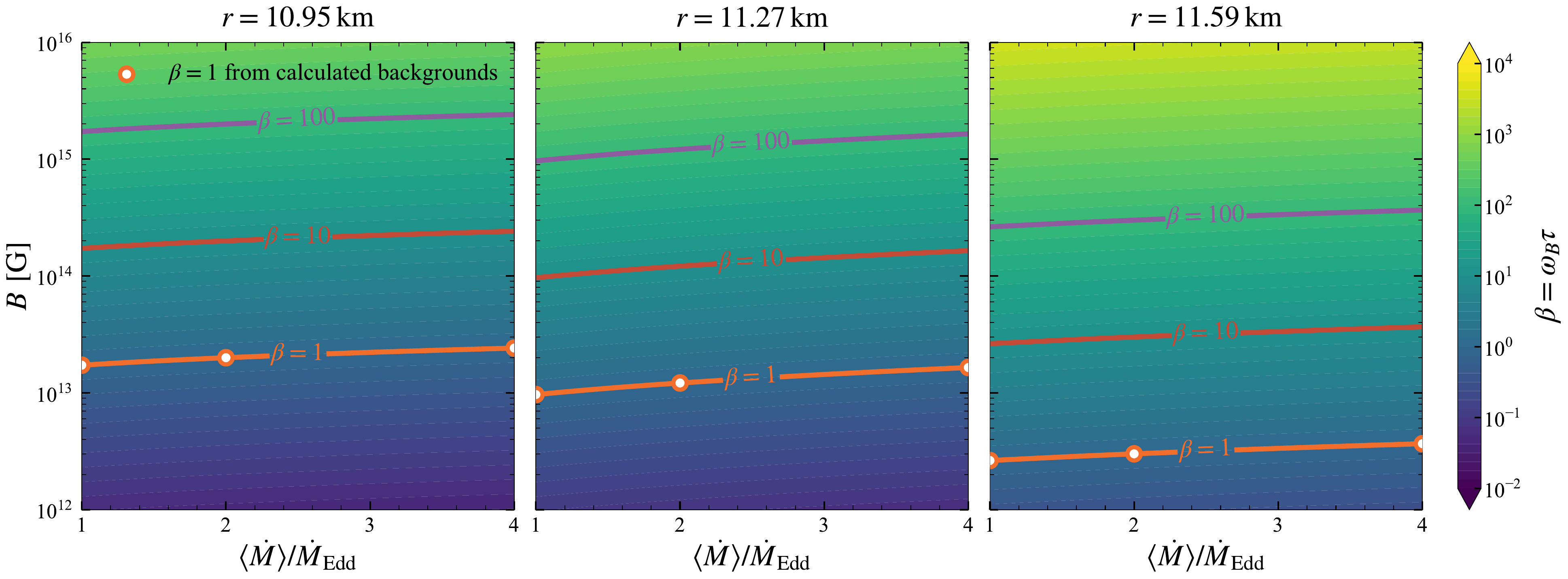}
    \caption{ Electron magnetization parameter in the $B$--$\langle\dot M\rangle$ plane at three representative crustal radii. The left, middle, and right panels correspond to $r=10.95$, $11.27$, and $11.59\,{\rm km}$, respectively. The color scale shows $\beta=\omega_{\rm B}\tau$, while the solid curves mark $\beta=1$, $10$, and $100$. Open circles indicate the magnetic-field strengths for which the directly calculated thermal backgrounds with $\langle\dot M\rangle/\dot M_{\rm Edd}=1$, $2$, and $4$ satisfy $\beta=1$. The magnetic-field dependence is evaluated using the classical scaling $\beta\propto B$ at fixed thermodynamic background.}
\label{fig:mag_parameter_2}
\end{figure*}
%%%%%%%%%%%%%%%%%%%%%%%%%%%%%%%%%%%

The combined dependence on magnetic-field strength, accretion rate, and crustal depth is shown in Fig.~\ref{fig:mag_parameter_2}. The transport coefficients perpendicular and parallel to the magnetic field satisfy
%----
\begin{equation}\label{eq:sm_kappa_ratio}
\frac{\kappa_\perp}{\kappa_\parallel}
=
\frac{1}{1+\beta^2} \,.
\end{equation}
%----
Accordingly, $\beta=1$, $10$, and $100$ correspond to $\kappa_\perp/\kappa_\parallel\simeq0.5$, $9.9\times10^{-3}$, and $10^{-4}$, respectively. The $\beta=1$ contour therefore provides a convenient measure of the onset of appreciable anisotropy, whereas $\beta=10$ and $100$ characterize progressively stronger suppression of transverse heat transport. At fixed radius, increasing $\langle\dot M\rangle$ shifts a given constant-$\beta$ contour toward larger $B$, because the hotter thermal background shortens the relaxation time. For example, at $r=10.95\,{\rm km}$ the field required to reach $\beta=1$ increases from approximately $1.7\times10^{13}\,{\rm G}$ at $\langle\dot M\rangle=\dot M_{\rm Edd}$ to $2.4\times10^{13}\,{\rm G}$ at $\langle\dot M\rangle=4\dot M_{\rm Edd}$.

The opposite trend occurs with crustal depth. Toward the outer crust, the smaller electron effective mass and the associated change in the scattering time lead to larger $\beta$ at a given $B$, so that the same degree of magnetization is reached at a weaker field. At $r=11.59\,{\rm km}$, for example, the $\beta=1$ transition occurs at only a few $\times10^{12}\,{\rm G}$. Across the accretion rates and depths considered here, we obtain
%----
\begin{equation}\label{eq:B_beta1_range}
B_{\beta=1}
\simeq
2.6\times10^{12}
-
2.4\times10^{13}\,{\rm G} \,.
\end{equation}
%----
We therefore identify $B\sim10^{13}\,{\rm G}$ as the characteristic transition between weakly and strongly anisotropic heat transport for the adopted crust model.

The parameter-space map in Fig.~\ref{fig:mag_parameter_2} relies on the classical scaling $\beta\propto B$, which assumes that the microscopic relaxation time is not itself modified by the magnetic field. This approximation is sufficient for identifying the transition between transport regimes. At sufficiently large fields, however, Landau quantization and magnetic modifications of the scattering rates may become important. The highest-field portions of the $\beta=10$ and $100$ contours should therefore be regarded as diagnostic extrapolations rather than precise microphysical predictions.

The diagnostics above treat $B$ as the local field strength used to identify the relevant transport regime. In the thermal calculations below, however, the magnetic field is spatially varying and is prescribed by a crust-confined toroidal $(\ell,m)=(2,2)$ geometry. We use the same geometry in both the weakly and strongly magnetized calculations so that the two regimes can be compared directly. Following \citet{Osborne:2019iph}, we write
%----
\begin{equation}\label{eq:sm_psi22}
\psi_{22}(r)
=
C[(r-R_{\rm IB})(r-R_{\rm OB})]^2 \,,
\end{equation}
%----
which gives
%----
\begin{align}\label{eq:sm_B22}
\mathbf{B}
&=
\frac{C}{4}
\sqrt{\frac{15}{2\pi}}
[(r-R_{\rm IB})(r-R_{\rm OB})]^2
\nonumber\\
&\quad\times
\left(
-\sin\theta\sin2\phi\,\mathbf{e}_{\theta}
-
\sin\theta\cos\theta\cos2\phi\,\mathbf{e}_{\phi}
\right) \,.
\end{align}
%----

We characterize the overall strength of this prescribed internal field by its value at the mid-crust reference point,
%----
\begin{equation}\label{eq:sm_Bref}
B_{\rm ref}
=
|\mathbf{B}|
\left(
\frac{R_{\rm IB}+R_{\rm OB}}{2} \,,
\frac{\pi}{2}\,,
\frac{\pi}{4}
\right) \,,
\end{equation}
%----
which fixes the normalization constant,
%----
\begin{equation}\label{eq:sm_C_normalization}
C
=
\frac{B_{\rm ref}}
{\displaystyle
\frac{1}{4}\sqrt{\frac{15}{2\pi}}
\frac{(R_{\rm OB}-R_{\rm IB})^4}{16}} \,.
\end{equation}
%----
The adopted radial dependence forces the field to vanish at both $R_{\rm IB}$ and $R_{\rm OB}$, thereby confining it to the computational crust. We emphasize that $B_{\rm ref}$ characterizes the prescribed internal toroidal field and should not be identified directly with the large-scale external dipole field inferred from spin evolution. This fixed magnetic geometry is used below to calculate the temperature response in the weakly and strongly magnetized regimes.

%---------------------------------------------------------------------
\subsection{Weakly Magnetized Regime}
\label{sec:Case_I}
%---------------------------------------------------------------------

For $\beta\ll1$, the magnetic modification of the conductive heat flux is small and can be treated perturbatively. We therefore expand the temperature and heat flux about the spherical background,
%----
\begin{equation}
T=T_0+\delta T \,,
\qquad
\mathbf F=\mathbf F_0+\delta\mathbf F \,,
\end{equation}
%----
and retain terms only to first order in both $\delta T$ and the magnetic correction.

With the sign convention adopted below for the Hall flux, the linearized heat-flux perturbation is
%----
\begin{equation}\label{eq:weak_deltaF}
\delta\mathbf F
=
-\left(\frac{\partial\kappa_0}{\partial T}\right)_{\rho,X_i}
\delta T\,\nabla T_0
-\kappa_0\nabla\delta T
-\kappa_0\varpi\tau\,
\mathbf B\times\nabla T_0 \,,
\end{equation}
%----
where
%----
\begin{equation}\label{eq:varpi_definition}
\varpi
\equiv
\frac{e}{m_e^\ast c}
=
\frac{\omega_{\rm B}}{B} \,.
\end{equation}
%----
Thus $\varpi\tau=\beta/B$. The magnetic-field amplitude is already contained in $\mathbf B$, or equivalently in the magnetic stream function introduced above; no additional factor of $B$ enters Eq.~(\ref{eq:weak_deltaF}). This makes the weak-field source term linear in $B_{\rm ref}$.

The steady energy equation is
%----
\begin{equation}
\nabla\cdot\mathbf F
=
Q_{\rm nuc}-Q_\nu \,.
\end{equation}
%----
Because the nuclear heating rate is independent of the local temperature in the present model, its magnetic perturbation vanishes. Hence
%----
\begin{equation}\label{eq:weak_perturbed_energy}
\nabla\cdot\delta\mathbf F
=
\frac{{\rm d}Q}{{\rm d}T}\delta T
=
-\frac{\partial Q_\nu}{\partial T}\delta T \,.
\end{equation}
%----
For the electron--ion bremsstrahlung emissivity adopted here,$Q_\nu\propto T^6$, so that
%----
\begin{equation}
\frac{{\rm d}Q}{{\rm d}T}
=
-\frac{6Q_\nu}{T_0} \,.
\end{equation}
%----

Because the background is spherically symmetric, we expand
%----
\begin{align}
\delta T(r,\theta,\phi)
&=
\sum_{\ell m}
\delta T_{\ell m}(r)
Y_{\ell m}(\theta,\phi) \,,
\\
\delta\mathbf F
&=
\sum_{\ell m}
\left[
U_{\ell m}(r)\hat{\mathbf r}Y_{\ell m}
+
V_{\ell m}(r)\nabla Y_{\ell m}
\right] \,.
\end{align}
%----
Substitution into Eq.~(\ref{eq:weak_perturbed_energy}) gives
%----
\begin{align}
\frac{{\rm d}U_{\ell m}}{{\rm d}r}
={}&
-\frac{2}{r}U_{\ell m}
+
\frac{\ell(\ell+1)}{r^2}V_{\ell m}
+
\frac{{\rm d}Q}{{\rm d}T}\delta T_{\ell m} \,,
\label{eq:weak_U_equation}
\\
\frac{{\rm d}\delta T_{\ell m}}{{\rm d}r}
={}&
-\frac{1}{\kappa_0}
\left(\frac{\partial\kappa_0}{\partial T}\right)_{\rho,X_i}
\frac{{\rm d}T_0}{{\rm d}r}
\delta T_{\ell m}
-\frac{U_{\ell m}}{\kappa_0} \,,
\label{eq:weak_deltaT_equation}
\end{align}
%----
together with
%----
\begin{equation}\label{eq:weak_V_relation}
V_{\ell m}
=
-\kappa_0\delta T_{\ell m}
+
\kappa_0\varpi\tau r
\frac{{\rm d}T_0}{{\rm d}r}
\psi_{\ell m} \,.
\end{equation}
%----

We retain the quadrupolar component with $\ell=m=2$ and impose the homogeneous radial boundary conditions
%----
\begin{equation}\label{eq:weak_boundary}
\delta T_{22}(R_{\rm IB})
=
\delta T_{22}(R_{\rm OB}) = 0 \,.
\end{equation}
%----
These conditions ensure that the nonaxisymmetric temperature perturbation is generated internally by magnetic heat transport; the accretion-column temperature enters only through the spherical background $T_0(r)$.

For the fiducial background, the weak-field solution satisfies $|\delta T_{22}|\propto B_{\rm ref}$. At $B_{\rm ref}=10^{12}\,{\rm G}$ we find
%----
\begin{equation}
\max
\left|
\frac{\delta T_{22}}{T_0}
\right|
\simeq
6.9\times10^{-4} \,.
\end{equation}
%----
This solution provides a direct weak-field benchmark for the multidimensional solver described below.

%---------------------------------------------------------------------
\subsection{Strongly Magnetized Regime}
\label{sec:Case_II}
%---------------------------------------------------------------------

When $\beta\gtrsim1$, transverse electron heat transport is appreciably suppressed and the weak-field expansion leading to Eq.~(\ref{eq:weak_deltaF}) is no longer valid. We therefore solve the full steady-state anisotropic heat-transport equation for the prescribed magnetic configuration. To fix the Hall-term convention, we write the conductive heat flux as
%----
\begin{align}
\mathbf F
={}&
-\kappa_\parallel
\mathbf b(\mathbf b\cdot\nabla T)
\nonumber\\
&-\kappa_\perp
\left[
\nabla T-\mathbf b(\mathbf b\cdot\nabla T)
\right]
-\kappa_{\rm H}\,
\mathbf b\times\nabla T \,,
\label{eq:strong_flux_vector}
\end{align}
%----
where $\mathbf b=\mathbf B/B$. Equivalently, $F_i=-\kappa_{ij}\nabla_jT$, with
%----
\begin{equation}\label{eq:strong_conductivity_tensor}
\kappa_{ij}
=
\kappa_\perp\delta_{ij}
+
(\kappa_\parallel-\kappa_\perp)b_i b_j
-
\kappa_{\rm H}\epsilon_{ijk}b_k \,,
\end{equation}
%----
where $\epsilon_{r\theta\phi}=+1$. The conductivity coefficients are
%----
\begin{equation}\label{eq:strong_kappa_components}
\kappa_\parallel=\kappa_0 \,,
\qquad
\kappa_\perp=\frac{\kappa_0}{1+\beta^2} \,,
\qquad
\kappa_{\rm H}=\frac{\beta\kappa_0}{1+\beta^2} \,.
\end{equation}
%----
The temperature field is then determined from
%----
\begin{equation}\label{eq:strong_heat_equation}
\nabla\cdot
\left[
\boldsymbol{\kappa}(T,\mathbf B)\cdot\nabla T
\right]
+
Q_{\rm nuc}(r)-Q_\nu(r,T)
=0 \,.
\end{equation}
%----
The transport coefficients are evaluated locally from the temperature and fixed crustal composition, and no expansion in either $\beta$ or $\delta T/T$ is imposed.

For the multidimensional calculation, we fix the temperature at both radial boundaries,
%----
\begin{align}
T(R_{\rm IB},\theta,\phi)
&=T_0(R_{\rm IB}) \,,
\label{eq:strong_inner_bc}\\
T(R_{\rm OB},\theta,\phi)
&=T_0(R_{\rm OB}) \,,
\label{eq:strong_outer_bc}
\end{align}
%----
together with periodicity in the azimuthal direction,
%----
\begin{equation}
T(r,\theta,\phi+2\pi)
= T(r,\theta,\phi) \,,
\end{equation}
%----
and regularity of the polar heat flux,
%----
\begin{equation}
\left[
\sin\theta\,F_\theta
\right]_{\theta=0,\pi}
=0 \,.
\end{equation}
%----
Equation~(\ref{eq:strong_heat_equation}) therefore determines the full three-dimensional temperature distribution $T(r,\theta,\phi)$ within the crust.

To separate the non-spherical thermal structure from its radial component, we define
%----
\begin{equation}\label{eq:strong_T_average}
\langle T\rangle(r)
=
\frac{1}{4\pi}
\int T(r,\theta,\phi)\,{\rm d}\Omega \,,
\end{equation}
%----
and
%----
\begin{equation}\label{eq:strong_deltaT_definition}
\delta T(r,\theta,\phi)
=
T(r,\theta,\phi)-\langle T\rangle(r) \,.
\end{equation}
%----
We project $\delta T$ onto the real $m=2$ quadrupolar basis,
%----
\begin{equation}\label{eq:real_Y22}
Y_{22}^{c,s}
=
\sqrt{\frac{15}{32\pi}}
\sin^2\theta
\left\{
\cos2\phi,\,
\sin2\phi
\right\} \,.
\end{equation}
%----
With this convention,
%----
\begin{equation}\label{eq:Y22_norm}
\int
\left(Y_{22}^{c,s}\right)^2
{\rm d}\Omega
=
\frac{1}{2} \,,
\end{equation}
%----
so the corresponding radial coefficients are
%----
\begin{align}
\delta T_{22}^{c,s}(r)
&=
\frac{
\displaystyle
\int
\delta T(r,\theta,\phi)
Y_{22}^{c,s}(\theta,\phi)\,
{\rm d}\Omega
}{
\displaystyle
\int
\left(Y_{22}^{c,s}\right)^2
{\rm d}\Omega
}
\nonumber\\
&=
2\int
\delta T(r,\theta,\phi)
Y_{22}^{c,s}(\theta,\phi)\,
{\rm d}\Omega \,.
\label{eq:strong_T22_components}
\end{align}
%----
For convenience, we also define the quadrupolar amplitude
%----
\begin{equation}\label{eq:strong_T22_amplitude}
|\delta T_{22}|
=
\left[
\left(\delta T_{22}^{c}\right)^2
+
\left(\delta T_{22}^{s}\right)^2
\right]^{1/2} \,.
\end{equation}
%----
This amplitude is used only as a diagnostic quantity; the signed coefficients $\delta T_{22}^{c}(r)$ and $\delta T_{22}^{s}(r)$ are retained separately when the thermal solution is coupled to the elastic-deformation equations. Because $\delta T_{22}$ is a harmonic expansion coefficient rather than a pointwise temperature variation, $\max(|\delta T_{22}|/\langle T_{0}\rangle]) $ need not be smaller than $\max[|\delta T|/\langle T\rangle]$.

The nonlinear equation is discretized in conservative finite-volume form on a three-dimensional spherical grid. We have checked the solution at several angular resolutions and find that the qualitative transition from the perturbative to the strongly anisotropic regime is robust, although the peak quadrupolar temperature coefficient retains some resolution dependence at the largest field strengths considered. The conductivity tensor and neutrino emissivity are updated iteratively from the current temperature field. At iteration $n$, the discretized transport equation is solved for a provisional field $T^{(n+1,\ast)}$, which is under-relaxed according to
%----
\begin{equation}
T^{(n+1)}
= \lambda T^{(n+1,\ast)} + (1-\lambda)T^{(n)} \,.
\end{equation}
%----
We monitor both the relative change between successive temperature iterates and the discrete steady-state residual,
%----
\begin{equation}\label{eq:strong_residual}
\eta
=
\frac{
\left|
\nabla\cdot
(\boldsymbol{\kappa}\cdot\nabla T)
+Q_{\rm nuc}-Q_\nu
\right|
}{
\left|
\nabla\cdot
(\boldsymbol{\kappa}\cdot\nabla T)
\right|
+
|Q_{\rm nuc}|+
|Q_\nu|
} \,.
\end{equation}
%----
The iteration is continued until the maximum relative temperature update falls below $10^{-7}$, while the residual is independently checked to ensure that the converged solution satisfies the discretized steady-state equation.

As an internal validation, we compare the full calculation with the perturbative solution in the weak-field limit. At $B_{\rm ref}=10^{12}\,{\rm G}$, the multidimensional calculation gives $\max(|\delta T_{22}|/T_0)=6.79\times10^{-4}$, compared with $6.87\times10^{-4}$ from the weak-field calculation, corresponding to agreement at the percent level. The difference remains below approximately $2\%$ at $B_{\rm ref}=5\times10^{12}\,{\rm G}$. This agreement provides a direct check that the nonlinear solver continuously recovers the perturbative solution as $\beta\rightarrow0$.

The departure from the weak-field behavior becomes apparent when the maximum magnetization approaches unity. At $B_{\rm ref}=10^{13}\,{\rm G}$, we obtain $\max\beta\simeq1.18$ and $\max(|\delta T_{22}|/T_0)\simeq6.22\times10^{-3}$, approximately
$9.5\%$ below the linear weak-field extrapolation. At higher field
strengths the full angular temperature distribution becomes
increasingly nonlinear. For $B_{\rm ref}=10^{14}\,{\rm G}$,
$\max[|\delta T|/\langle T\rangle]\simeq0.243$ and
$\max(|\delta T_{22}|/T_0)\simeq8.45\times10^{-2}$, while at the
largest field considered here,
$B_{\rm ref}=3\times10^{14}\,{\rm G}$, these quantities increase to
approximately $0.490$ and $0.220$, respectively, with
$\max\beta\simeq15.7$. We therefore restrict the quantitative
analysis to $B_{\rm ref}\leq3\times10^{14}\,{\rm G}$, beyond which
the effects of extreme transport anisotropy, angular resolution, and
magnetic modifications of the microscopic transport coefficients
become increasingly important.

%---------------------------------------------------------------------
\begin{table*}[t]
\caption{Summary of the fixed-field thermal calculations for the quadrupolar toroidal magnetic configuration. The second column lists the maximum local magnetization parameter. The last two columns give the maximum local fractional temperature variation and the maximum $(\ell,m)=(2,2)$ temperature coefficient, respectively. }
\begin{ruledtabular}
\begin{tabular}{cccc}
$B_{\rm ref}\,[{\rm G}]$
&
$\max\beta$
&
$\max[|\delta T|/\langle T\rangle]$
&
$\max(|\delta T_{22}|/T_0)$
\\
\hline
$10^{11}$          & $1.18\times10^{-2}$ & $2.84\times10^{-5}$ & $6.97\times10^{-5}$ \\
$10^{12}$          & $1.18\times10^{-1}$ & $4.15\times10^{-4}$ & $6.79\times10^{-4}$ \\
$5\times10^{12}$   & $5.91\times10^{-1}$ & $5.43\times10^{-3}$ & $3.38\times10^{-3}$ \\
$10^{13}$          & $1.18$               & $1.58\times10^{-2}$ & $6.22\times10^{-3}$ \\
$3\times10^{13}$   & $3.30$               & $5.31\times10^{-2}$ & $1.92\times10^{-2}$ \\
$10^{14}$          & $7.38$               & $2.43\times10^{-1}$ & $8.45\times10^{-2}$ \\
$1.5\times10^{14}$ & $9.23$               & $3.52\times10^{-1}$ & $1.40\times10^{-1}$ \\
$2\times10^{14}$   & $11.21$              & $4.20\times10^{-1}$ & $1.87\times10^{-1}$ \\
$3\times10^{14}$   & $15.74$              & $4.90\times10^{-1}$ & $2.20\times10^{-1}$ \\
\end{tabular}
\end{ruledtabular}
\label{tab:fixed_b_scan}
\end{table*}
%---------------------------------------------------------------------

Table~\ref{tab:fixed_b_scan} shows that the weak-field scaling is maintained while $\max\beta<1$. Around $B_{\rm ref}\sim10^{13}\,{\rm G}$, $\max\beta$ becomes of order unity and the quadrupolar temperature response begins to depart from the perturbative extrapolation. For $B_{\rm ref}\gtrsim10^{14}\,{\rm G}$, both the local thermal contrast and the quadrupolar component become large, confirming that the crust has entered a strongly anisotropic, nonlinear transport regime.

We also evaluate the local Coulomb coupling parameter using the full multidimensional temperature field,
%---
\begin{equation}
\Gamma_{\rm Coul}(r,\theta,\phi)
=
\frac{Z^2e^2}
{a_i k_{\rm B}T(r,\theta,\phi)} \,.
\end{equation}
%---
For the strongest model, $B_{\rm ref}=3\times10^{14}\,{\rm G}$, the high-density inner crust remains crystalline, while localized melting occurs in the lower-density outer crust, where the hottest regions fall below the canonical one-component-plasma melting threshold $\Gamma_{\rm m}=175$. The first such regions appear at $\rho\simeq10^{11}\,{\rm g\,cm^{-3}}$. Because the present elastic calculation assumes a spherically stratified solid crust and does not incorporate an angle-dependent solid--liquid interface, the quadrupole obtained for the strongest-field model should be regarded as an upper-end estimate associated with the full-solid approximation.

%---------------------------------------------------------------------
\section{Elastic deformation calculation}
\label{sec:sm_elastic}
%---------------------------------------------------------------------

We calculate the elastic response of the crust following \citet{Ushomirsky:2000ax}. In the Cowling approximation, we introduce the total Eulerian stress perturbation
%----
\begin{equation}\label{eq:total_stress}
\delta\tau_{ij}
=
-\delta P\,g_{ij}
+t_{ij} \,,
\end{equation}
%----
where $t_{ij}$ denotes the shear-stress perturbation. The perturbed force-balance equation is then
%----
\begin{equation}\label{eq:elastic_force_balance}
\nabla^j\delta\tau_{ij}
=
\delta\rho\,g\,\nabla_i r \,,
\end{equation}
%----
where $g>0$ is the magnitude of the Newtonian gravitational acceleration. For an isotropic elastic solid,
%----
\begin{equation} \label{eq:shear_stress}
t_{ij}
=
\mu
\left(
\nabla_i\xi_j+\nabla_j\xi_i
-\frac{2}{3}g_{ij}\nabla_k\xi^k
\right) \,,
\end{equation}
%----
where $\xi^i$ is the displacement vector. We adopt
%----
\begin{equation}
\mu
=
0.1194\,
\frac{n_{\rm ion}(Ze)^2}{a} \,,
\qquad
a=
\left(
\frac{3}{4\pi n_{\rm ion}}
\right)^{1/3} \,.
\end{equation}
%----

For a polar perturbation, the displacement is expanded as
%----
\begin{equation}\label{eq:displacement}
\xi^i
=
\xi_r(r)Y_{\ell m}\nabla^i r
+
\frac{r}{\lambda}
\xi_\perp(r)\nabla^iY_{\ell m} \,,
\qquad
\lambda\equiv\sqrt{\ell(\ell+1)} \,.
\end{equation}
%----
The shear stress is decomposed into the corresponding tensor spherical harmonics \cite{Ushomirsky:2000ax},
%----
\begin{align}\label{eq:shear-stress_tensor}
t_{ij}
={}&
t_{rr}Y_{\ell m}
\left(
\nabla_i r\nabla_j r
-\frac{1}{2}e_{ij}
\right)
+t_{r\perp}f_{ij}
\nonumber\\
&+
t_{\Lambda}
\left(
\Lambda_{ij}
+\frac{1}{2}Y_{\ell m}e_{ij}
\right) \,,
\end{align}
%----
where
%----
\begin{align}
e_{ij}
&=
g_{ij}-\nabla_i r\nabla_j r \,,
\\
f_{ij}
&=
\frac{r}{\lambda}
\left(
\nabla_i r\nabla_jY_{\ell m}
+
\nabla_j r\nabla_iY_{\ell m}
\right) \,,
\\
\Lambda_{ij}
&=
\left(
\frac{r}{\lambda}
\right)^2
\nabla_i\nabla_jY_{\ell m}
+
\frac{1}{\lambda}f_{ij} \,.
\end{align}
%----
The radial component entering the traction condition is the total stress perturbation,
%----
\begin{equation}\label{eq:total_traction_components}
\delta\tau_{rr}
=
-\delta P+t_{rr} \,,
\qquad
\delta\tau_{r\perp}=t_{r\perp} \,,
\end{equation}
%----
so that the pressure perturbation is explicitly included.

The perturbed continuity equation gives
%----
\begin{equation}
\delta\rho
=
-\rho\frac{{\rm d}\xi_r}{{\rm d}r}
-
\left(
\frac{2\rho}{r}
+\frac{{\rm d}\rho}{{\rm d}r}
\right)\xi_r
+
\lambda\rho\frac{\xi_\perp}{r} \,.
\end{equation}
%----

Following \citet{Ushomirsky:2000ax}, we introduce the dimensionless variables
%----
\begin{align}
z_1
&=
\frac{\xi_r}{r} \,,
\\
z_2
&=
\frac{\Delta\tau_{rr}}{P}
=
\frac{\delta\tau_{rr}}{P}
-
z_1\frac{{\rm d}\ln P}{{\rm d}\ln r} \,,
\\
z_3
&=
\frac{\xi_\perp}{\lambda r} \,,
\\
z_4
&=
\frac{\Delta\tau_{r\perp}}{\lambda P}
=
\frac{\delta\tau_{r\perp}}{\lambda P} \,,
\end{align}
%----
where $\Delta$ denotes a Lagrangian perturbation. These variables satisfy
%----
\begin{align}
\frac{{\rm d}z_1}{{\rm d}\ln r}
={}&
-
\left(
1+2\frac{\alpha_2}{\alpha_3}
-\frac{\alpha_4}{\alpha_3}
\right)z_1
+
\frac{z_2}{\alpha_3}
+
\ell(\ell+1)
\frac{\alpha_2}{\alpha_3}z_3
+
\frac{\Delta S}{\alpha_3} \,,
\label{eq:bc_p1}
\\
\frac{{\rm d}z_2}{{\rm d}\ln r}
={}&
\left(
UV-4V
+12\Gamma\frac{\alpha_1}{\alpha_3}
-4\frac{\alpha_1\alpha_4}{\alpha_3}
\right)z_1
+
\left(
V-4\frac{\alpha_1}{\alpha_3}
\right)z_2
\nonumber\\
&+
\ell(\ell+1)
\left(
V-6\Gamma\frac{\alpha_1}{\alpha_3}
\right)z_3
+
\ell(\ell+1)z_4
-
4\frac{\alpha_1}{\alpha_3}\Delta S \,,
\label{eq:bc_p2}
\\
\frac{{\rm d}z_3}{{\rm d}\ln r}
={}&
-z_1+\frac{z_4}{\alpha_1} \,,
\label{eq:bc_p3}
\\
\frac{{\rm d}z_4}{{\rm d}\ln r}
={}&
\left(
V-6\Gamma\frac{\alpha_1}{\alpha_3}
+2\frac{\alpha_1\alpha_4}{\alpha_3}
\right)z_1
-
\frac{\alpha_2}{\alpha_3}z_2
\nonumber\\
&+
\frac{2}{\alpha_3}
\left\{
[2\ell(\ell+1)-1]\alpha_1\alpha_2
+
2[\ell(\ell+1)-1]\alpha_1^2
\right\}z_3
\nonumber\\
&+
(V-3)z_4
+
2\frac{\alpha_1}{\alpha_3}\Delta S \,.
\label{eq:bc_p4}
\end{align}
%----
Here
%----
\begin{equation}
U
\equiv
\frac{{\rm d}\ln g}{{\rm d}\ln r}+2 \,,
\qquad
V
\equiv
\frac{\rho gr}{P}
=
-\frac{{\rm d}\ln P}{{\rm d}\ln r} \,,
\end{equation}
%----
and
%----
\begin{equation}\label{eq:elastic_coefficients}
\alpha_1
\equiv
\frac{\mu}{P} \,,
\qquad
\alpha_2
\equiv
\Gamma-\frac{2\mu}{3P} \,,
\qquad
\alpha_3
\equiv
\Gamma+\frac{4\mu}{3P} \,.
\end{equation}
%----

For a temperature-driven deformation,
%----
\begin{equation}
\Gamma
=
\left.
\frac{\partial\ln P}{\partial\ln\rho}
\right|_T \,,
\qquad
\alpha_4
=
\left.
\frac{\partial\ln P}{\partial\ln T}
\right|_\rho
\frac{{\rm d}\ln T_0}{{\rm d}\ln r} \,.
\end{equation}
%----
For each real quadrupolar component, the thermal source is
%----
\begin{equation}\label{eq:elastic_thermal_source}
\Delta S_{22}^{c,s}(r)
=
\left.
\frac{\partial\ln P}{\partial\ln T}
\right|_\rho
\frac{\delta T_{22}^{c,s}(r)}{T_0(r)} \,,
\end{equation}
%----
where $T_0(r)$ is retained as the spherical reference background about which the elastic response is linearized. We note that subtracting the angular mean $\langle T\rangle$ rather than $T_0$ does not change the signed $(\ell,m)=(2,2)$ temperature coefficient, because the difference $\langle T\rangle-T_0$ is spherically symmetric. At the largest fields, the nonlinear thermal solution also modifies the angularly averaged temperature profile. As a sensitivity test, we replace $T_0(r)$ in the normalization of the thermal source by $\langle T\rangle(r)$. For $B_{\rm ref}=3\times10^{14}\,{\rm G}$, after restricting the elastic calculation to the connected crystalline region identified above, this replacement reduces $|Q_{22}|$ from $3.18\times10^{37}$ to $2.78\times10^{37}\,{\rm g\,cm^2}$, corresponding to a change of approximately $13\%$. We therefore regard the strongest-field quadrupole as a linear-response estimate with a residual systematic uncertainty associated with the choice of the spherical thermal reference state.

The cosine and sine components are propagated independently through Eqs.~(\ref{eq:bc_p1})--(\ref{eq:bc_p4}). Since the elastic system is linear in $\Delta S$, we calculate $Q_{22}^{c}$ and $Q_{22}^{s}$ separately and combine them only after solving the two problems,
%----
\begin{equation}\label{eq:Q22_amplitude}
|Q_{22}|
=
\left[
(Q_{22}^{c})^2+
(Q_{22}^{s})^2
\right]^{1/2} \,.
\end{equation}
%----

At the crust--core interface, where we neglect a density discontinuity, the traction boundary conditions are
%----
\begin{equation}
z_2=Vz_1 \,,
\qquad
z_4=0 \,.
\end{equation}
%----
At the crust--ocean interface they become
%----
\begin{equation}
z_2
=
V\frac{\rho_l}{\rho_s}z_1 \,,
\qquad
z_4=0 \,,
\end{equation}
%----
where $\rho_s$ and $\rho_l$ are the densities on the solid and liquid sides of the interface. We use $(\rho_s-\rho_l)/\rho_s\simeq5\times10^{-5}$, corresponding to $\rho_l/\rho_s \simeq 0.99995$.

Once Eqs.~(\ref{eq:bc_p1})--(\ref{eq:bc_p4}) have been solved, the mass multipole is evaluated from
%----
\begin{align}\label{eq:Q_2}
Q_{\ell m}
=
-\int_{r_{\rm bot}}^{r_{\rm top}}
\frac{1}{V}
\bigg\{&
\ell(\ell+1)z_4
-
2\alpha_1
\left[
2\frac{{\rm d}z_1}{{\rm d}\ln r}
+
\ell(\ell+1)z_3
\right]
\nonumber\\
&+
(\ell+4-U)
(z_2-Vz_1)
\bigg\}
r^{\ell+2}\rho\,{\rm d}r \,,
\end{align}
%----
where $r_{\rm bot}$ and $r_{\rm top}$ denote the bottom and top of the elastic crust.

Using the signed temperature profiles obtained from the thermal calculation, we solve the elastic equations independently for the cosine and sine quadrupolar components. The mass quadrupole increases substantially with $B_{\rm ref}$, reaching $|Q_{22}|\simeq1.0\times10^{36}\,{\rm g\,cm^2}$ at $10^{12}\,{\rm G}$, $8.6\times10^{36}\,{\rm g\,cm^2}$ at $10^{13}\,{\rm G}$, and $7.8\times10^{37}\,{\rm g\,cm^2}$ at $10^{14}\,{\rm G}$. For the strongest field considered here, $B_{\rm ref}=3\times10^{14}\,{\rm G}$, the direct calculation using the nonlinear signed temperature profile gives
%----
\begin{equation}
|Q_{22}|
=
3.22\times10^{38}\,{\rm g\,cm^2} \,,
\qquad
\epsilon
=
4.17\times10^{-7} \,.
\end{equation}
%----
At this field strength, the multidimensional temperature solution indicates localized regions in the low-density outer crust where the temperature exceeds the melting temperature, while the deeper crust, which provides most of the elastic support, remains solid. Since the present one-dimensional elastic treatment cannot represent an angle-dependent melting boundary, we retain the nominal radial extent of the elastic crust. The resulting quadrupole should therefore be interpreted as an upper-end estimate of the thermo-elastic deformation.

The corresponding cosine component dominates, while the sine component is consistent with numerical zero for the adopted magnetic phase convention. The peak quadrupolar temperature coefficient shows a moderate dependence on angular resolution in the strongly magnetized regime, whereas the integrated mass quadrupole is more stable against grid refinement. For the same elastic-domain prescription, the three tested thermal grids give $|Q_{22}|=(3.12-3.22)\times10^{38}\,\mathrm{g\,cm^2}$, corresponding to a spread of approximately $3.5\%$. This comparison tests the numerical sensitivity of the fixed-domain calculation to the adopted thermal grids; it does not quantify the systematic uncertainty associated with changes in the solid--liquid interfaces. We quote $|Q_{22}|\simeq3.2\times10^{38}\,\mathrm{g\,cm^2}$ fixed-domain calculation.

We evaluate the linear elastic response on the solid domain defined by the spherical thermal background $T_0(r)$. This domain is held fixed when the signed quadrupolar components of the nonlinear three-dimensional temperature field are used as the thermal source. In particular, the solid--liquid interfaces are not updated to account for local melting in the nonlinear thermal solution. The largest-field quadrupole should therefore be interpreted as a fixed-domain thermo-elastic estimate; whether a comparable deformation can be sustained when melting-induced changes in the elastic domain are included remains to be established. The grid-resolution tests reported above assess numerical sensitivity within the prescribed domain and do not quantify the systematic uncertainty associated with such changes. The detectability estimates in the main text are consequently conditional on the star sustaining a quadrupole of comparable magnitude.

%---------------------------------------------------------------------
\section{Spin evolution calculation}
\label{sec:spin_evolution}
%---------------------------------------------------------------------

We follow the framework of \citet{Ho:2020vtg} to model the spin evolution of ULXPs through the ejector, propeller, and accretor phases. An important distinction in the present calculation is that the mass quadrupole considered here is generated by accretion-induced temperature asymmetries. We therefore assume that the corresponding thermal mountain is absent before sustained surface accretion begins. Consequently, $Q_{22}=0$ during the ejector and ideal-propeller phases, whereas the quadrupole develops after the system enters the accretor phase. Hereafter, $Q_{22}$ denotes the amplitude of the $(\ell,m)=(2,2)$ mass quadrupole.

The magnetic-dipole spin-down torque is
%----
\begin{equation}
\label{eq:EM}
N_{\rm EM}
=
-\frac{2\mu^2\Omega^3\sin^2\theta}{3c^3}
=
-K_{\rm EM} I\Omega^3 \,,
\end{equation}
%----
where
%----
\begin{equation}
K_{\rm EM}
\equiv
\frac{2\mu^2\sin^2\theta}{3c^3 I} \,,
\end{equation}
%----
$\Omega=2\pi/P$ is the stellar spin angular frequency and $\theta$ is the angle between the rotation and magnetic axes. We denote the surface polar dipole field by $B_{\rm p}$ and adopt the magnetic dipole-moment convention
%----
\begin{equation}
\mu=\frac{1}{2}B_{\rm p}R_*^3 \,.
\end{equation}
%----
We assume an orthogonal rotator, $\sin^2\theta=1$, and adopt the fiducial NS parameters
$M_*=1.4\,M_\odot$ \,,
$R_*=10^6\,{\rm cm}$\,, and
$I=10^{45}\,{\rm g\,cm^2}$ \,.

The spin-down torque associated with GW emission is \citep{Ushomirsky:2000ax}
%----
\begin{equation}\label{eq:GW}
N_{\rm GW}
=
-\frac{256\pi}{75}
\frac{GQ_{22}^{2}}{c^{5}}
\Omega^{5} \,.
\end{equation}
%----
In our fiducial evolutionary model, this torque is absent before the onset of sustained surface accretion because the accretion-induced thermal quadrupole has not yet developed.

According to standard accretion theory \citep{1979ApJ...234..296G, 1987A&A...183..257W}, sustained surface accretion cannot occur while the magnetospheric radius lies outside the light cylinder. The light-cylinder radius is
%----
\begin{equation}\label{eq:r_lc}
r_{\rm lc}
=
\frac{c}{\Omega}
=
48\,{\rm km}
\left(
\frac{P}{1\,{\rm ms}}
\right) \,.
\end{equation}
%----
The magnetospheric radius is written as \citep{1972A&A....21....1P}
%----
\begin{equation}\label{eq:r_m}
r_{\rm m}
=
\xi r_A
=
\xi
\left(
\frac{\mu^4}
{8GM_*\dot M_{\rm in}^{\,2}}
\right)^{1/7} \,,
\end{equation}
%----
where $\dot M_{\rm in}$ is the effective mass-inflow rate relevant to the magnetospheric interaction, $\xi$ is a dimensionless parameter typically in the range $0.5$--$1.4$ \citep{1996ApJ...465L.111W}, and $r_A$ is the Alfv\'en radius, defined as the characteristic radius at which the ram pressure of the inflowing material becomes comparable to the magnetic pressure.

When
%----
\begin{equation}
r_{\rm m}>r_{\rm lc} \,,
\end{equation}
%----
the system is in the ejector phase. Since the thermal mountain has not yet formed, we set $Q_{22}=0$, and the spin evolution is governed by magnetic-dipole radiation,
%----
\begin{equation}
\label{eq:evolution_1}
\frac{{\rm d}\Omega}{{\rm d}t}
=
-K_{\rm EM}\Omega^3 \,.
\end{equation}
%----
The initial angular frequency is
%----
\begin{equation}
\Omega_0=\frac{2\pi}{P_0} \,,
\end{equation}
%----
and we adopt $P_0=1\,{\rm ms}$ for the representative evolutionary track.

As the NS spins down, $r_{\rm lc}$ increases until the light cylinder expands beyond the magnetosphere, $r_{\rm lc}=r_{\rm m}$, marking the transition into the propeller phase. The corotation radius is
%----
\begin{equation}\label{eq:r_co}
r_{\rm co}
=
\left(
\frac{GM_*}{\Omega^2}
\right)^{1/3}
=
17\,{\rm km}
\left(
\frac{M_*}{1.4\,M_\odot}
\right)^{1/3}
\left(
\frac{P}{1\,{\rm ms}}
\right)^{2/3} \,.
\end{equation}
%----
In the ideal-propeller regime, $r_{\rm co}<r_{\rm m}<r_{\rm lc}$, the centrifugal barrier inhibits sustained accretion onto the stellar surface. We therefore continue to set $Q_{22}=0$ and describe the spin evolution using
%----
\begin{equation}\label{eq:evolution_2}
I\frac{{\rm d}\Omega}{{\rm d}t}
=
-\dot M_{\rm in} r_{\rm m}^{2}
\left[
\Omega-\Omega_K(r_{\rm m})
\right] \,,
\end{equation}
%----
where
%----
\begin{equation}
\Omega_K(r_{\rm m})
=
\left(
\frac{GM_*}{r_{\rm m}^{3}}
\right)^{1/2} \,.
\end{equation}
%----
Within the piecewise prescription adopted here, the vacuum magnetic-dipole torque is used in the ejector phase and is neglected once the inflowing matter penetrates inside the light cylinder.

For fixed $B_{\rm p}$ and $\dot M_{\rm in}$, both $r_{\rm m}$ and $\Omega_K(r_{\rm m})$ are constant, and Eq.~(\ref{eq:evolution_2}) drives $\Omega$ exponentially toward $\Omega_K(r_{\rm m})$. Exact equality is therefore reached only asymptotically in this simplified torque prescription. To assign a finite transition time into the near-equilibrium accretor state, we introduce the fastness parameter
%----
\begin{equation}\label{eq:fastness}
\omega_{\rm s}
\equiv
\frac{\Omega}{\Omega_K(r_{\rm m})} \,,
\end{equation}
%----
and define the practical onset of the accretor phase by
%----
\begin{equation}\label{eq:accretor_criterion}
\omega_{\rm s}
\leq
1+\epsilon_{\rm acc} \,.
\end{equation}
%----
We adopt $\epsilon_{\rm acc}=0.03$ in the fiducial calculation. This parameter is only a numerical criterion used to define a finite transition time and should not be confused with the stellar ellipticity $\epsilon$. The nominal material-torque equilibrium corresponds to $\omega_{\rm s}=1$, with
%----
\begin{equation}
P_{\rm eq}
\equiv
P_K(r_{\rm m})
=
\frac{2\pi}{\Omega_K(r_{\rm m})} \,.
\end{equation}
%----
Thus, the adopted value of $\epsilon_{\rm acc}$ shifts the practical transition period by only a few percent from the nominal equilibrium period.

Once sustained surface accretion begins, an accretion column can form and the nonaxisymmetric thermal deformation develops. We parameterize the evolution of the corresponding mass quadrupole as
%----
\begin{equation}\label{eq:Q22_evolution}
\frac{{\rm d}Q_{22}}{{\rm d}t}
=
\frac{Q_{22}^{\rm ss}-Q_{22}}{\tau_Q} \,,
\end{equation}
%----
where $Q_{22}^{\rm ss}$ is the steady-state quadrupole obtained from the thermal-transport and elastic-deformation calculations, and $\tau_Q$ characterizes the thermal-mountain build-up timescale. Since the thermal calculations presented in this work are performed in the steady-state limit, $\tau_Q$ is not determined self-consistently. We therefore treat it as a sensitivity parameter and adopt
%----
\begin{equation}
\tau_Q=10\,{\rm yr}
\end{equation}
%----
for the fiducial evolutionary calculation.

With $Q_{22}(t_{\rm acc})=0$, Eq.~(\ref{eq:Q22_evolution}) gives
%----
\begin{equation}\label{eq:Q22_solution}
Q_{22}(t)
=
Q_{22}^{\rm ss}
\left[
1-
\exp\left(
-\frac{t-t_{\rm acc}}{\tau_Q}
\right)
\right] \,,
\qquad
t\geq t_{\rm acc} \,,
\end{equation}
%----
where $t_{\rm acc}$ denotes the adopted onset of the accretor phase. We have verified that varying $\tau_Q$ over the range $1$--$100\,{\rm yr}$ changes the transient quadrupole build-up but has a negligible effect on the subsequent spin evolution. In particular, $P_{\rm acc}$ is independent of $\tau_Q$, because the quadrupole is assumed to develop only after $t_{\rm acc}$.

The spin evolution after the onset of surface accretion is then governed by
%----
\begin{equation}\label{eq:evolution_accretor}
I\frac{{\rm d}\Omega}{{\rm d}t}
=
-\dot M_{\rm in} r_{\rm m}^{2}
\left[
\Omega-\Omega_K(r_{\rm m})
\right]
+
N_{\rm GW} \,,
\end{equation}
%----
where $N_{\rm GW}$ is evaluated using the time-dependent quadrupole in Eq.~(\ref{eq:Q22_solution}). The material torque vanishes at $\Omega=\Omega_K(r_{\rm m})$ and becomes positive when $\Omega<\Omega_K(r_{\rm m})$. The additional GW spin-down torque therefore shifts the exact torque balance slightly below $\Omega_K(r_{\rm m})$; for the fiducial parameters adopted here, this shift is negligible.

For the representative evolutionary track shown in Fig.~\ref{fig:evolution}, we adopt
%----
\begin{equation}
B_{\rm p}=10^{12}\,{\rm G} \,,
\qquad
\dot M_{\rm in}
=
8.5\times10^{19}\,{\rm g\,s^{-1}} \,,
\qquad
\xi=0.5 \,,
\end{equation}
%----
together with $P_0=1\,{\rm ms}$.

We distinguish the effective inflow rate $\dot M_{\rm in}$ used in the spin-evolution calculation from the long-term crustal accretion rate $\langle\dot M\rangle$ entering the thermal calculation. The former determines the magnetospheric radius and material torque, whereas the latter controls the secular deep-crustal heating and the background thermal structure. In the fiducial thermal calculation we adopt $\langle\dot M\rangle=\dot M_{\rm Edd}$, while the illustrative spin-evolution track uses the substantially larger $\dot M_{\rm in}=8.5\times10^{19}\,{\rm g\,s^{-1}}$ to represent a strongly super-Eddington inflow. We do not identify these two rates. A fully self-consistent mapping between them would require modeling the mass-transfer history, duty cycle, super-Eddington mass loss, and the fraction of the inflowing material ultimately incorporated into the NS crust, which is beyond the scope of the present work.

For the GW torque after the onset of accretion, we adopt the steady-state quadrupole obtained from our fiducial thermal-mountain model with $\langle\dot M\rangle=\dot M_{\rm Edd}$ and $B_{\rm ref}=3\times10^{14}\,{\rm G}$,
%----
\begin{equation}
Q_{22}^{\rm ss}
=
3.22\times10^{38}\,{\rm g\,cm^2} \,.
\end{equation}
%----
This value is used as a representative strong-field thermal deformation and is not recalculated along the illustrative spin-evolution track.

We likewise distinguish the two magnetic-field scales entering the calculation. The external polar dipole field $B_{\rm p}=10^{12}\,{\rm G}$ determines the magnetospheric evolution, whereas $B_{\rm ref}=3\times10^{14}\,{\rm G}$ characterizes the prescribed crust-confined field responsible for anisotropic heat transport and the thermal mountain. No equality between $B_{\rm p}$ and $B_{\rm ref}$ is assumed.

The adopted spin-evolution parameters give
%----
\begin{equation}
r_{\rm m}
\simeq
124\,{\rm km} \,,
\end{equation}
%----
and therefore
%----
\begin{equation}
P_K(r_{\rm m})
=
\frac{2\pi}{\Omega_K(r_{\rm m})}
\simeq
20.16\,{\rm ms} \,.
\end{equation}
%----
Using $\epsilon_{\rm acc}=0.03$, the practical transition into the near-equilibrium accretor state occurs at
%----
\begin{equation}
P_{\rm acc}
\simeq
\frac{P_K(r_{\rm m})}
{1+\epsilon_{\rm acc}}
\simeq
19.6\,{\rm ms} \,.
\end{equation}
%----
The numerical integration gives an ejector-to-propeller transition at approximately $3.7\times10^{2}\,{\rm yr}$ and a propeller-to-accretor transition at approximately $1.35\times10^{4}\,{\rm yr}$. Because the accretion-induced quadrupole is set to zero before the onset of sustained surface accretion, $P_{\rm acc}$ is determined by the magnetic and accretion parameters and is unaffected by the adopted value of $Q_{22}^{\rm ss}$. The updated quadrupole modifies only the post-accretion GW torque and produces a negligible change in the fiducial spin evolution.

For completeness, we estimate the characteristic gravitational energy-release rate during the evolution as
%----
\begin{equation}
\label{eq:evolution_3}
L_{\rm grav}
\simeq
\frac{GM_*\dot M_{\rm in}}{r}
=
1.2\times10^{36}\,
{\rm erg\,s^{-1}}
\left(
\frac{M_*}{1.4\,M_\odot}
\right)
\left(
\frac{\dot M_{\rm in}}
{10^{-10}\,M_\odot\,{\rm yr^{-1}}}
\right)
\left(
\frac{10\,{\rm km}}{r}
\right) \,.
\end{equation}
%----
For illustrative purposes, we take
%----
\begin{equation}
r=
\begin{cases}
r_{\rm lc}, & {\rm ejector}\,,\\
r_{\rm m},  & {\rm propeller}\,,\\
R_*,        & {\rm accretor}\,.
\end{cases}
\end{equation}
%----
The corresponding evolution is shown in the lower panel of Fig.~\ref{fig:evolution}. The values assigned to the ejector and propeller phases should be interpreted only as characteristic gravitational-power scales, because matter does not undergo sustained accretion onto the stellar surface in these phases. Neither phase is therefore required to be ultraluminous. The relevant requirement for the present work is that the system becomes super-Eddington after entering the accretor phase, which is satisfied for the adopted parameters.

Equation~(\ref{eq:evolution_3}) neglects the detailed radiative efficiency, mass loss, and geometric beaming associated with super-Eddington accretion \citep{Middleton:2016qss, 2017MNRAS.471L..71M, King:2019jvg, Lasota:2023xef}. It should therefore be regarded as an order-of-magnitude estimate of the available gravitational power rather than as a detailed prediction of the observed isotropic-equivalent X-ray luminosity.

\end{document}